\documentclass[12pt]{article}
\usepackage{epsf,amssymb,amsmath,psfrag}
\usepackage[dvips]{graphicx}
\usepackage{color,colordvi}
\def\b#1{\Blue{#1}}
\catcode`\@=11
\def \thesection {\arabic{section}.}

\def \be  {\begin{equation}}
\def \ee  {\end{equation}}
\def \ba  {\begin{eqnarray}}
\def \ea  {\end{eqnarray}}
\def \baa {\begin{eqnarray*}}
\def \eaa {\end{eqnarray*}}
\def \bb  {\begin {thebibliography} }
\def \eb  {\end{thebibliography}}
\def \lab #1 {\label{#1}}

\newcommand\re[1]{(\ref{#1})}

\def \matrix #1 {\left(\begin{array}{cc} #1 \end{array}\right)}

\def \tr {\mathop{\rm tr}\nolimits}

\def \Re {\mathop{\rm Re}\nolimits}

\def \e  {\mathop{\rm e}\nolimits}
\newcommand\lr[1]{{\left({#1}\right)}}

\newcommand \vev [1] {\langle{#1}\rangle}

\newcommand{\as}{\ifmmode\alpha_{\rm s}\else{$\alpha_{\rm s}$}\fi}
\newcommand{\asbar}{\ifmmode\bar{\alpha}_{\rm s}\else{$\bar{\alpha}_{\rm s}$}\fi}

\newcommand{\ft}[2]{{\textstyle\frac{#1}{#2}}}

\font\cmss=cmss12 
\def\inbar{\,\vrule height1.5ex width.4pt depth0pt}
\def\IC{\relax\hbox{$\inbar\kern-.3em{\rm C}$}}
\def\IZ{\relax{\hbox{\cmss Z\kern-.4em Z}}}
\def\IR{{\hbox{{\rm I}\kern-.2em\hbox{\rm R}}}}

\def\IP{{\hbox{{\rm I}\kern-.2em\hbox{\rm P}}}}
\def\II{\hbox{{1}\kern-.25em\hbox{l}}}

\def\numberbysection{\@addtoreset{equation}{section}
                     \def\theequation{\thesection\arabic{equation}}}
\numberbysection

\newbox\lett\newdimen\lheight\newdimen\lwidth
\def\ontop#1#2{\setbox\lett=\hbox{#2}\lheight\ht\lett
\multiply\lheight by 12 \divide\lheight by 10\relax%
\lwidth\wd\lett \multiply\lwidth by 8 \divide\lwidth by 10\relax #2\kern-\lwidth%
\raise\lheight\hbox{{$\scriptstyle #1$}}\kern.1ex}

\begin{document}

\begin{titlepage}
\begin{flushright}
\begin{tabular}{l}
LPT--Orsay--06--99 \\
hep-th/0612247
\end{tabular}
\end{flushright}

\vskip3cm

\centerline{\large \bf Anomalous dimensions of high-spin operators beyond the
leading order}

\vspace{1cm}

\centerline{\sc B. Basso, G.P. Korchemsky}

\vspace{10mm}

\centerline{\it Laboratoire de Physique Th\'eorique\footnote{Unit\'e
                    Mixte de Recherche du CNRS (UMR 8627).},
                    Universit\'e de Paris XI}
\centerline{\it 91405 Orsay C\'edex, France}

\def\thefootnote{\fnsymbol{footnote}}%
\vspace{1cm}

\centerline{\bf Abstract}

\vspace{5mm}

Anomalous dimensions of Wilson operators with large Lorentz spin scale
logarithmically with the spin. Recent multi-loop QCD calculations of twist-two
anomalous dimensions revealed the existence of interesting structure of the
subleading corrections suppressed by powers of the Lorentz spin. We argue that
this structure is a manifestation of the `self-tuning' property of the multi-loop
anomalous dimensions -- in a conformal gauge theory, the anomalous dimension of
Wilson operators is a function of their conformal spin which is modified in
higher loops by an amount proportional to the anomalous dimension. Making use of
the parity property of this function and incorporating the beta-function
contribution, we demonstrate the existence of (infinite number of) relations
between subleading corrections to the twist-two anomalous dimensions in QCD and
its supersymmetric extensions. They imply that the subleading corrections to the
anomalous dimensions suppressed by odd powers of the conformal spin can be
expressed in terms of the lower-loops corrections suppressed by smaller even
powers of the spin. We show that these relations hold true in QCD to all loops in
the large $\beta_0$ limit. In the $\mathcal{N}=4$ SYM theory, we employ the
AdS/CFT correspondence to argue that the same relations survive in the strong
coupling regime for higher-twist scalar operators dual to a folded string
rotating on the AdS${}_3\times$S${}^1$.

\vspace*{20mm}

\centerline{ \textsl{to the memory of Alexander Nikolaevich Vasil'ev} }

\end{titlepage}

\setcounter{footnote} 0

%{\small \tableofcontents}

\newpage

\section{Introduction}

Success of QCD as the theory of strong interactions relies in part on the
possibility to predict the scale dependence of various physical observables. In
the most prominent example of deeply inelastic scattering, this dependence is
governed by anomalous dimensions of composite Wilson operators that arise in the
operator product expansion of conserved currents. The anomalous dimensions depend
on the coupling constant as well as on the quantum numbers of operators (Lorentz
spin, isotopic charge) and their properties reflect the symmetries of the
underlying gauge theory. In the maximally supersymmetric $\mathcal{N}=4$
Yang-Mills theory, the gauge/string correspondence allows one to relate the
anomalous dimensions at strong coupling to the energies of strings propagating on
the AdS${}_5\times$S${}^5$ background~\cite{Mal97}. In the case of QCD, due to
lack of a similar dual stringy description, one can only calculate the anomalous
dimensions at weak coupling as series in the coupling constant.

At present, the most advanced QCD calculations of anomalous dimensions have been
performed to three-loop accuracy for the Wilson operators of twist
two~\cite{MVV04,VMV04}. The resulting expressions for multi-loop anomalous
dimensions are complicated functions of the Lorentz spin $N$ carried by the
Wilson operators but they simplify significantly for large $N$. In this limit,
for $N\gg 1$, the anomalous dimensions scale logarithmically $\gamma(N) \sim \ln
N$ with subleading corrections running in inverse powers of $N$~\cite{Kor88}. The
explicit three-loop calculation~\cite{MVV04,VMV04} revealed the existence of
interesting structure of the subleading corrections. It turned out that, to
three-loop accuracy, correction $\sim 1/N$ to the twist-two anomalous dimension
can be expressed in terms of the leading $\sim N^0$ correction evaluated to two
loops only (MVV relation). Recently, it was shown in Refs.~\cite{DMS} that this
relation can be explained by modifying the evolution (renormalization group)
equation in such a way \cite{DKT95} as to preserve the reciprocity relation
between its solutions (see Eq.~\re{GL} below).

In this paper, we shall argue that the MVV relation is a manifestation of the
following `self-tuning' property of the multi-loop anomalous
dimensions~\cite{BKM06} -- in a conformal gauge theory, the anomalous dimension
is a function of the conformal spin of the Wilson operator which is modified at
higher loops by an amount proportional to the anomalous dimension. Making use of
the parity property of this function and incorporating the beta-function
contribution, we shall demonstrate the existence of (infinite number of) MVV like
relations between subleading corrections to the twist-two anomalous dimensions at
large $N$ in QCD and its supersymmetric extensions. We shall show that subleading
corrections to the anomalous dimensions suppressed by odd powers of $1/N$ are not
independent and can be expressed in terms of coefficients accompanying smaller
{\it even} powers of $1/N$. % to {\it less} number of loops.
In the $\mathcal{N}=4$ SYM theory, we shall employ the AdS/CFT correspondence to
argue that the same relations survive in the strong coupling regime for scalar
operators of higher twist.

The anomalous dimensions of twist-two operators govern the scale dependence of
nonperturbative parton (quark, gluon) distribution functions in the deeply
inelastic scattering (DIS) \cite{GL72a,AP77,D77}. They are related to the parton
space-like (${\scriptstyle S}$) splitting functions $P_S(z)$ through the Mellin
transform
\be\label{z-repr}
\gamma_S(N) = - \int_0^1 dz \, z^{N-1} P_S(z)\,.
\ee
The function $P_S(z)$ defined in this way has a clear physical meaning. It
defines the probability for a parton to split into another parton carrying the
momentum fraction $z$.\footnote{The splitting functions cease to have a
probabilistic interpretation at higher twists~\cite{JS82}.} One encounters
another set of nonperturbative parton fragmentation functions in hadron
production in the $\e^+\e^--$annihilation which is just the cross-channel of the
DIS process (see review \cite{CTEQ} and references therein). The scale dependence
of the fragmentation functions is governed by the time-like (${\scriptstyle T}$)
splitting functions $P_T(z)$ whose moments provide yet another %twist-two
anomalous dimensions $\gamma_T(N)=- \int_0^1 dz \, z^{N-1} P_T(z)$. In
distinction with the parton distribution functions, the fragmentation functions
do not admit the operator product expansion and, as a consequence, their
anomalous dimensions $\gamma_T(N)$ are {\it not} related to composite local
Wilson operators. The two functions are related however by the crossing symmetry.
This suggests to write down the following reciprocity relations between the
space-like, $P_S(z)$, and time-like, $P_T(z)$, splitting functions
\be\label{GL}
    P_T(z) = P_S(z)\,, \qquad
    P_T(z) = -z\, P_S(1/z)\,,
\ee
the so-called Gribov-Lipatov~\cite{GL72} and Drell-Levy-Yan~\cite{DLY69}
relations. Here in the first relation the two functions are equated within their
physical region $0\le z <1$, while in the second relation the space-like
splitting function is analytically continued through $z=1$. In general, the two
relations in \re{GL} are independent on each other. However, if they were
fulfilled simultaneously, one would obtain a `strong' reciprocity relation for
$P(z)=P_T(z) = P_S(z)$
\be\label{self}
P(z) = -z\, P(1/z)\,.
\ee
It is known \cite{CFP80,FP80,FKL81,SV96,SV01} that this relation is satisfied to
one loop only, while both relations in \re{GL} are violated to two loops.
\footnote{Notice that the second relation in \re{GL} still holds true to two
loops for the so-called physical evolution kernels governing the scale dependence
of the DIS structure functions and the hadron fragmentation
functions~\cite{BRN00}.}

Another set of relations for the twist-two anomalous dimensions comes from a
space-time picture of parton (quark, gluon) splitting in the space-like
(distribution) and time-like (fragmentation) processes. In the latter case,
imposing the condition of the angular ordering for successive partonic splittings
one obtains the following remarkable functional equation for the leading
asymptotic behaviour of the time-like gluon anomalous dimension $\gamma_T(N)$ at
{\it small} $N$~\cite{Mue83}
\be\label{M}
\gamma_T(N) = \gamma_S(N-\gamma_T(N))\,,\qquad \gamma_S(N) = -\frac{\alpha_s
N_c}{\pi {N} } + \ldots\,,
\ee
where ellipses stand for subleading terms as $N\to 0$ and
$\alpha_s/N^2=g_s^2/(4\pi N^2)=\mbox{fixed}$. The relation \re{M} leads to a
quadratic equation for $\gamma_T(N)$ whose solution resums the leading
corrections $\sim \alpha_s^n/N^{2n-1}$ to $\gamma_T(N)$  for all $n\ge 1$. It has
been proposed in Refs.~\cite{DKT95,DMS} that a relation similar to \re{M}
should also hold at {\it large} $N$ %
\footnote{Due to different conventions for the anomalous dimension (see
Eq.~\re{RG} below), $\gamma_\sigma(N)$ differs from similar expression in
Ref.~\cite{DMS} by the overall factor $(-1/2)$.}
\be\label{DMS}
\gamma_\sigma(N) = f\lr{N-\ft12 \sigma \gamma_\sigma(N)}\,,
\ee
where $\sigma=-$ for the space-like and $\sigma=+$ for the time-like anomalous
dimensions, $\gamma_-(N)=\gamma_S(N)$ and $\gamma_+(N)=\gamma_T(N)$. The function
$f(N)$ is assumed to be the same for $\sigma=\pm$ and, most importantly, its
large $N$ expansion takes the form
\be\label{DMS-f}
f(N) = A\left[\psi(N+1)+\gamma_{\rm\scriptscriptstyle E}\right] + B + 0\cdot
N^{-1}+ \mathcal{O}(N^{-2})\,,
\ee
where $\psi(x)=d\ln\Gamma(x)/dx$ and $\gamma_{\rm\scriptscriptstyle E}$ is the
Euler constant. Then, the relation \re{DMS} leads to the following expression for
the twist-two (nonsinglet quark and diagonal elements of the singlet mixing
matrix) anomalous dimensions at large~$N$
\be\label{DMS-as}
\gamma_\sigma (N) = A \ln\bar N + B + C_\sigma N^{-1} \ln \bar N +
\lr{D_\sigma+\ft12 A} N^{-1} +\mathcal{O}(N^{-2}),
\ee
with $\bar N= N\e^{\gamma_{\rm\scriptscriptstyle E}}$. Here the leading term is
related to the universal cusp anomalous dimension~\cite{Kor88}
\be
A= 2\Gamma_{\rm cusp}(\alpha_s)\,,
\ee
the $B-$coefficient is the same for the space-like and time-like distributions
but it depends on the quantum numbers of partons (flavour, spin projection). From
\re{DMS} and \re{DMS-f} one finds the remaining coefficients as~\cite{DMS}
\be\label{DMS-ACD}
C^{\rm \scriptscriptstyle (MVV)}_\sigma = -\ft12 {\sigma} A^2(\alpha_s)\,,\qquad
D_\sigma = -\ft12 {\sigma} A(\alpha_s) B(\alpha_s)\,.
\ee
The expression for $C_\sigma$ is in agreement with explicit three-loop QCD result
for the space-like~\cite{MVV04,VMV04} and (nonsinglet) time-like anomalous
dimensions~\cite{MMV06}, while for $D_\sigma$ the mismatch is proportional to the
beta-function
\be\label{D-exact}
D_\sigma^{\rm \scriptscriptstyle (MVV)} = -\ft12 {\sigma} A(\alpha_s) B(\alpha_s)
-\ft12 A(\alpha_s)\beta(\alpha_s)\,.
\ee
As was emphasized in Refs.~\cite{MVV04,DMS}, this calls for a further structural
explanation of subleading $1/N$ corrections to the anomalous dimensions
\re{DMS-as} at large $N$ and, more generally, for a better understanding of the
origin of the functional relations \re{M} and \re{DMS}.

As we will see momentarily, for the space-like anomalous dimensions
$\gamma_S(N)$, the relation \re{DMS} is ultimately tied to conformal invariance
of a classical four-dimensional Yang-Mills theory. Moreover, the conformal
invariance allows one to extend the relation \re{DMS} to anomalous dimensions of
the so-called quasipartonic operators~\cite{BFLK85} of arbitrary twist $L$. A
distinguished feature of these operators is that they are built from exactly $L$
fundamental fields of twist $1$ and from an arbitrary number of covariant
derivatives projected onto the light-cone. The twist-two operators correspond to
$L=2$.

The classical Yang-Mills Lagrangian is invariant under conformal transformations
but this symmetry is broken on the quantum level unless the beta-function
vanishes to all loops (see  review \cite{BKM03} and references therein). If the
conformal symmetry was exact (as in $\mathcal{N}=4$ SYM theory), the
quasipartonic operators could be classified according to representations of the
collinear $SL(2;\mathbb{R})$ subgroup of the full $SO(2,4)$ conformal
group~\cite{O81}. Conformal invariance ensures that the operators belonging to
different $SL(2;\mathbb{R})$ multiplets cannot mix under renormalization and
their anomalous dimension, $\gamma_S(N)$, depends on the conformal
$SL(2;\mathbb{R})$ spin of the multiplet. By definition, the conformal spin of
the quasipartonic operator of twist $L$ equals $j=\ft12(N+\ell)$ with $N$ and
$\ell$ being its Lorentz spin and the scaling dimension, respectively. The former
is protected from perturbative corrections, whereas the latter equals $N+L$ to
the lowest order in the coupling constant and receives anomalous contribution
$\gamma_S(N)$ at higher orders. As a result, the conformal spin of quasipartonic
operators gets modified in higher loops as~\cite{M93,BKM04,BKM06}
\be\label{J-ren}
j_{\rm bare} = N+\ft12L \quad \mapsto \quad  j = N+\ft12L+\ft12 \gamma_{S}(N)\,.
\ee
Then, the conformal symmetry implies that the anomalous dimension of the
quasipartonic operator is a function of $j$, or equivalently
\be\label{f-beta=0}
\gamma_S(N) = f^{(\beta=0)}\lr{N+\ft12\gamma_S(N)}\,,
\ee
where the scaling function $f(N)$ depends on twist $L$ and other quantum numbers
of the operator. Here the superscript indicates that the function $f^{(\beta=0)}$
is  defined in gauge theory with vanishing beta-function. The relation
\re{f-beta=0} coincides with the first relation in \re{DMS} in the space-like
case $\sigma=-$.

In gauge theory with nonvanishing beta-function, the relation \re{f-beta=0}
should be modified to incorporate the additional conformal symmetry breaking
corrections. The latter are known to be renormalization scheme dependent. The
analysis can be simplified by renormalizing the Wilson operators in the
dimensional regularization scheme (DREG) with $d=4-2\varepsilon$.%
\footnote{In SYM theories, to preserve the supersymmetry, one uses the
dimensional reduction scheme (DRED) instead.} In this case, the coupling constant
acquires a nonvanishing dimension and the beta-function in
$(4-2\varepsilon)-$dimensional gauge theory is given by
\be
\beta_\varepsilon(\alpha_s)= -2\varepsilon + \beta(\alpha_s)\,.
\ee
A crucial observation is that $\beta_\varepsilon(\alpha_s)$ vanishes at
$\varepsilon_{\rm cr}=\beta(\alpha_s)/2$ and, therefore, the gauge theory is
conformal in $d_{\rm cr}=4-2\varepsilon_{\rm cr}$ dimensions. This opens up the
possibility to calculate the anomalous dimensions within the
$\varepsilon-$expansion using the conformal field theory technique as pioneered
in Refs.~\cite{VPK,V}. As before, the quasipartonic operators can be classified
according to representation of the $SL(2;\mathbb{R})$ collinear subgroup of the
conformal group in $d_{\rm cr}-$dimensions and their anomalous dimension is a
function of the conformal spin $j_{\rm cr}=\ft12(N+\ell_{\rm cr})$. The only
difference as compared with the $d=4$ case is that for $\varepsilon_{\rm cr}\neq
0$ the scaling dimensions of fundamental fields get shifted by
$(-\varepsilon_{\rm cr})$ so that the scaling dimension of the quasipartonic
operator built from $L$ fields  takes the form $\ell_{\rm
cr}=N+L-L\varepsilon_{\rm cr}+\gamma_S(N)$. Then, the conformal invariance in
$d_{\rm cr}-$dimensions implies that the anomalous dimension of twist$-L$
operators in $d=4-$dimensional gauge theory is a function of the conformal spin
$j_{\rm cr}$
\be\label{f-def}
\gamma_S(N) = f^{(\beta\neq 0)}\left(N+\ft12\gamma_S(N) -\ft14 L
\beta(\alpha_s)\right)\,.
\ee
It is important to stress that, in distinction with $f^{(\beta=0)}(N)$, the
function $f^{(\beta\neq 0)}(N)$ depends on the renormalization scheme. Invoking
the same arguments that led to \re{DMS}, one can extend the relation \re{f-def}
to the time-like anomalous dimension
\be\label{T-def}
\gamma_T(N) = f^{(\beta\neq 0)}\left(N-\ft12\gamma_T(N) -\ft14 L
\beta(\alpha_s)\right)\,.
\ee
The equations \re{f-def} and \re{T-def} generalize the relations \re{DMS}
proposed in \cite{DKT95,DMS}. They relate the space-like and time-like anomalous
dimension of twist $L$ to the same function $f^{(\beta\neq 0)}(N)$ and
incorporate the beta-function contribution.

Conformal symmetry allows one to relate the space-like anomalous dimension
$\gamma_S(N)$ to yet another function $f(N)$ but it does not tell us much about
the properties of the latter. Therefore, to make use of the functional relation
\re{f-def} it has to be supplemented with additional relation for the scaling
function $f(N)$. As a hint, let us consider the twist-two anomalous dimension
\re{DMS-as} and examine the first few terms in the large $N$ expansion of $f(N)$,
Eq.~\re{DMS-f}. Their substitution into \re{f-def} and \re{T-def} yields (for
$L=2$) the large $N$ expansion of the space-like and time-like anomalous
dimensions \re{DMS-as} with the $D-$coefficient matching the exact three-loop
result \re{D-exact}. Notice that it is due to zero value of the coefficient in
front of $N^{-1}$ in the right-hand side of \re{DMS-f} that the $C-$ and
$D-$coefficients parameterizing subleading $\sim N^{-1}$ corrections to the
anomalous dimension can be expressed in terms of the leading coefficients,
Eqs.~\re{DMS-ACD} and \re{D-exact}. The question arises whether similar relations
also exist for the subleading $\sim N^{-n}$ corrections (for $n\ge 2$). We shall
argue that the answer is positive and it relies on the following remarkable
property of the function $f(N)$: the large $N$ expansion of the function $f(N)$
only runs in {\em integer negative} powers of the parameter
\be\label{J2}
{J}^2 = N (N+1)\,,
\ee
which has the meaning of a `bare' quadratic Casimir of the collinear
$SL(2;\mathbb{R})$ group. For twist-two operators, the same property can be
expressed as (with $\bar J= J \e^{\gamma_{\rm\scriptscriptstyle E}}$)
\be\label{prop}
f(N) = \Gamma_{\rm cusp}(\alpha_s) \ln {\bar J}^2  + f^{(0)} + f^{(1)}/{J}^2+
f^{(2)}/ {J}^4 + \mathcal{O}(1/{J}^6)\,,
\ee
where  $f^{(k)}$ (with $k\ge 0$) are given by series in $\alpha_s$ with the
coefficients depending on $\ln {\bar J}$. The relation \re{prop} implies that
each term of the asymptotic series \re{prop} is invariant (modulo contribution
from $\ln {\bar J}$) under $J \to -J$, or equivalently $N\to -1-N$. In what
follows we shall refer to \re{prop} as a {\it parity preserving} asymptotic series. %
\footnote{The same property would not be manifest if one replaced $J^2$ by its
expression \re{J2} and re-expanded $f(N)$ in powers of $1/N$. The resulting
series for $f(N)$ involves both even {and} odd powers of $1/N$ and its first
three terms match \re{DMS-f} for $f^{(0)}=B$.} We will show that the relations
\re{f-def}, \re{T-def} and \re{prop} are in a perfect agreement with available
two- and three-loop results for various anomalous dimensions in QCD and in
supersymmetric Yang-Mills theories.

The relation \re{prop} has a number of important consequences.  Firstly, going
back from the $N-$space to the $z-$representation, $f(N) = - \int_0^1 dz\,z^{N-1}
P_f(z)$, one finds that \re{prop} is translated into the reciprocity condition
\re{self} for the splitting function $P_f(z)$. Secondly, substituting \re{prop}
into \re{f-def} and \re{T-def}, one obtains that the coefficients in front of
${N}^{-2n}$ and $N^{-2n-1}$  in the large $N$ expansion of  the twist-two
anomalous dimensions are related to {\em the same} coefficients $f^{(m)}$ with
$0\le m \le n$. Excluding the latter, one gets an infinite set of relations
between the former (odd and even) coefficients, including the relations
\re{DMS-ACD} and \re{D-exact} for $n=0$. Moreover, re-expanding the anomalous
dimensions at large $N$ in inverse powers of $J=\sqrt{N(N+1)}$, one finds that
corrections suppressed by odd powers of $J$ can be expressed in terms of the
coefficients accompanying smaller even powers of $J$ to less number of loops.

The relations \re{f-def} and \re{prop} are universal properties of the twist-two
operators in gauge theories -- they hold true both in QCD and in supersymmetric
Yang-Mills theories with even number of supercharges $\mathcal{N}=0,2,4$, to
three loops at least. In $\mathcal{N}=1$ theory the function $f(N)$ develops an
additional square-root singularity $\sim\sqrt{1/J^2}$ due to degeneracy of the
parity-even and parity-odd anomalous dimensions in the large $N$ limit. Finally,
the relation \re{prop} can be also tested to all loops: In QCD, we will determine
the function $f(N)$ in the so-called large $\beta_0-$limit and show that it
verifies \re{prop} indeed. In the $\mathcal{N}=4$ theory, we will apply the
gauge/string correspondence to evaluate $f(N)$ in the strong coupling regime for
the special class of scalar operators of higher twist dual to a folded string
rotating on the AdS${}_3\times$S${}^1$ part of the target space of the type IIB
string theory \cite{GKP02,FT03}.

The paper is organized as follows. In Sect.~2, we apply well-known one-loop
expressions for various twist-two anomalous dimensions in QCD and in SYM theories
to justify the parity preserving relation \re{prop} and to establish its
connection with the reciprocity relation \re{self}. In Sect.~3, we use explicit
results for two-loop and three-loop anomalous dimensions available in the
literature to verify the relations \re{f-def}, \re{T-def} and \re{prop} beyond
the leading order. Then, we make use of the parity preserving property \re{prop}
to find the relations between different terms in the large $N$ expansion of the
anomalous dimensions analogous to \re{DMS-ACD} and \re{D-exact}. In Sect.~4, we
argue that the property \re{prop} holds true in QCD for twist-two anomalous
dimensions to all loops in the large $\beta_0$ limit. In Sect.~5, we employ the
AdS/CFT correspondence to determine the function $f(N)$ in the planar
$\mathcal{N}=4$ theory in the strong coupling regime for scalar operators of
higher twist. Section~6 contains concluding remarks. Some details of our
calculations are presented in the Appendix.

\section{Symmetries of twist-two anomalous dimensions}

Let us start with the definition of the space-like anomalous dimensions in a
generic four-dimensi\-onal Yang-Mills theory with the $SU(N_c)$ gauge group. In
general, the Wilson operators with the same Lorentz spin $N$ mix under
renormalization and verify the Callan-Symanzik equation
\be\label{RG}
\lr{ \frac{\partial}{\partial \ln\mu}+ \beta(\lambda)\frac{\partial}{\partial \ln
\lambda}} \mathcal{O}_N^{(a)}(0) = - \gamma^{ab}(N) \mathcal{O}_N^{(b)}(0)\,,
\ee
where the coupling constant $\lambda$ is defined slightly differently in QCD and
in SYM theory
\be\label{couplage}
\lambda^{\rm \scriptscriptstyle (QCD)} = \frac{g_s^2}{8\pi^2} \equiv
\frac{\alpha_s}{2\pi}\,,\qquad \lambda^{\rm \scriptscriptstyle (SYM)} = \frac{g^2
N_c}{8\pi^2}\,.
\ee
Here the mixing matrix is given by a series in the coupling constant
\be\label{gamma-exp}
\gamma(N) = \lambda \gamma_0(N) + \lambda^2 \gamma_1(N) + \ldots\,,
\ee
with $\gamma_k(N)$ being matrices, and the beta-function is defined as
\be
\frac{d \ln \lambda}{d\ln\mu} = \beta(\lambda) = -
\beta_0\lambda -  \beta_1\lambda^2 + \mathcal{O}(\lambda^3)\,.
\ee
The beta-function coefficients are given in QCD by
\be\label{beta-QCD}
\beta_0^{\rm \scriptscriptstyle (QCD)}=\frac{11}3\,{N_c}-\frac23\,{n_f}\,,\qquad
\beta_1^{\rm \scriptscriptstyle (QCD)}={\frac {17}{3}}\,N_c^{2}-\frac53\,{N_c}{n_f}-
{C_F} {n_f}
\ee
with $n_f$ the number of quark flavours and $C_F=(N_c^2-1)/(2N_c)$ the quadratic
Casimir in the fundamental representation of the $SU(N_c)$ group. In the SYM
theory with $\mathcal{N}$ supercharges one has
\be\label{beta-N}
\beta_0^{\rm \scriptscriptstyle (SYM)} = (4- \mathcal{N})  \,,\qquad
\beta_1^{\rm \scriptscriptstyle (SYM)} = ( 2- \mathcal{N})( 4-
\mathcal{N})\,.
\ee
For $\mathcal{N}=2$ the exact beta-function is given by the one-loop expression
while for $\mathcal{N}=4$ it vanishes to all loops and the corresponding gauge
theory is conformal.

\subsection{Nonsinglet anomalous dimensions in QCD}

Let us first consider nonsinglet twist-two quark operators in QCD. They take the
form
\be\label{O-ns}
{\mathcal{O}_N^{\rm (ns)}(0)} = {\bar q_{\alpha,i}\,
\frac{}{}T^a_{ij}\,\Gamma_{\alpha\beta} D_+^{N-1} q_{\beta,j}(0)}\,,
\ee
where $q_{\beta\!,j}$ denotes the quark field component carrying the flavour $j$
and the spinor index $\beta$. Also, $D_+=\partial_+ + i g A_+$ is the light-cone
component of the covariant derivative. The matrices $T^a_{ij}$ stand for the
generators of the flavour $SU(n_f)$ group while the spinor matrices
$\Gamma_{\alpha\beta}$ serve to select the so-called ``good'' (twist-one)
components of quark and antiquark fields either with a definite helicity, or with
a transverse polarization~\cite{BKM03}. A distinguished feature of the operators
\re{O-ns} is that they can not mix under renormalization with other operators but
only with the operators $\partial_+^n {\mathcal{O}_{N-n}^{{}_{\rm (ns)}}(0)}$
containing total derivatives. This mixing can be avoided by going over from the
operators ${\mathcal{O}_N^{\rm \scriptscriptstyle (ns)}(0)}$ to their forward
matrix elements with respect to some reference state $\vev{\mathcal{O}_N^{\rm
(ns)}(0)}$.

Depending on the choice of the $\Gamma-$matrices in \re{O-ns} one can distinguish
three different twist-two operators: unpolarized $(\Gamma=\gamma_+)$,
longitudinally polarized $(\Gamma=\gamma_+\gamma_5)$ and
transversity $(\Gamma=\gamma_+\gamma_\perp)$ operators.%
\footnote{In the case of transversity, the operators \re{O-ns} evolve
autonomously even for flavour singlet quark distributions. The same holds true
for the linearly polarized gluon distribution~\cite{V98}.} To the lowest order in
the coupling constant their anomalous dimensions are given by well-known
expressions~\cite{BFLK85}
\be\label{anom0}
\gamma_0(N) = C_F \left[4\psi(N+1)+4\gamma_{\rm\scriptscriptstyle
E}-3-\frac{1+\eta}{N(N+1)} \right],
\ee
where $\eta=1$ for the unpolarized and longitudinally polarized operators and
$\eta=-1$ for the transversity operator. Here $\psi(x) = d\ln \Gamma(x)/dx$ and
$\gamma_{\rm\scriptscriptstyle E}=-\psi(1)$ is the Euler constant. In the
momentum fraction representation \re{z-repr}, the corresponding splitting
functions read
\be\label{split}
P_0(z) = C_F\left[\frac{4z}{(1-z)_+} + 3 \delta(1-z) + (1-z)(1+\eta) \right],
\ee
where the `+' distribution is defined in a standard way
\be
\int_0^1 dz \,\varphi(z) \frac1{(1-z)_+} = \int_0^1
dz\frac{\varphi(z)-\varphi(1)}{1-z}
\ee
with $\varphi(z)$ being an arbitrary test function.

As was already mentioned in the Introduction, to one-loop order the space-like
and time-like splitting functions satisfy the reciprocity relations \re{GL}.
Indeed, one verifies that for $z\neq 1$ the splitting function $P_{0}(z)$,
Eq.~\re{split}, fulfills the relation \re{self}. The same property in the
$N-$space finds its manifestation in the asymptotic behaviour of the anomalous
dimension $\gamma_0(N)$ for large $N$. In this limit, the anomalous dimension
\re{anom0} scales as $\gamma_0(N) = 4C_F\ln N+\ldots$ with subleading corrections
running in powers of $1/N$. However, changing the expansion parameter from $N$ to
$J^2=N(N+1)$ and expanding the anomalous dimensions in inverse powers of $1/J$
one finds that the subleading corrections
run in {\em even} powers of $1/J$ only\\[0mm]
\be\label{gamma0-as}
\gamma_0(N) = C_F \left[2\,\ln \bar J^2 -3 - \lr{\ft13+\eta}
J^{-2}-\ft2{15}\,{J}^{-4}+{\ft {16}{315}}\,{J}^{-6}+\mathcal{O} \left( {J}^{-8}
\right)\right],
\ee\\[0mm]
where $\bar J=J\e^{\gamma{\rm\scriptscriptstyle E}}$. As follows from \re{f-def},
the functions $\gamma_S(N)$ and $f(N)$ coincide to the lowest order in the
coupling constant, $f(N) = \lambda \gamma_0(N)+ \mathcal{O}(\lambda^2)$,  but
differ from each other starting from two loops. Matching \re{gamma0-as} into
\re{prop}, one can identify one-loop contribution to the cusp anomalous dimension
$\Gamma_{\rm cusp}(\alpha_s)$ and to the coefficients $f^{(n)}$.

Each term of the expansion \re{gamma0-as} is invariant under substitution $J\to
-J$, or equivalently $N\to -1-N$, whereas the exact expression \re{anom0} is not
\be\label{cot}
\gamma_0(N) - \gamma_0(-N-1) = -4 C_F\, \pi\cot(\pi N)\,.
\ee
Still, the two relations, Eqs.~\re{gamma0-as} and \re{cot}, are consistent with
each other -- the one-loop anomalous dimension \re{anom0} has (Regge) poles at
integer negative $N$ and expansion \re{gamma0-as} is only well-defined for $\Re N
> 0$. Regge singularities of $\gamma_0(N)$ lead to a factorial growth of the expansion
coefficients in the right-hand side of \re{gamma0-as} (they scale as $\sim (-1)^n
{(2n)!}/{(n(2\pi J)^{2n})}$ for $n \gg 1$). \footnote{The asymptotic series
\re{gamma0-as} can be resummed using the Borel transform $ \gamma_0(N) =
\int_0^\infty dx \, \e^{-x/\alpha} \tilde\gamma (x)$, with $\alpha=1/N$ being the
expansion parameter. Indeed, for $z=\e^{-x}$ this relation coincides with the
definition of the moments \re{z-repr} provided that
$\tilde\gamma(x)=P_{0}(\e^{-x})$.}

Let us now show that the validity of the reciprocity relation \re{self} for the
splitting function and the absence of odd powers of $1/J$ in the asymptotic
expansion of the anomalous dimension \re{z-repr} are in the one-to-one
correspondence with each other. To this end, one changes the integration variable
in \re{z-repr} as $z= \e^{-x/j}$ with $j=N+\ft12=\lr{J^2+\ft14}^{1/2}$ and
evaluates derivative  $\gamma'(N)= \partial_N \gamma(N)$ as
\be\label{sample}
\gamma'(N) = \frac1{j^2} \int_0^\infty dx \,  \e^{-x} \left[ x \e^{x/(2j)}
P(\e^{-x/j})\right].
\ee
Expanding the integrand at large $j$ and integrating term-by-term one can obtain
an asymptotic expansion of $\gamma'(N)$ in powers of $1/j$. If asymptotic
expansion of $\gamma(N)$ run in even powers of $1/J$, or equivalently $1/j$, then
similar expansion of the derivative $\gamma'(N)$ would run in odd powers of $1/j$
leading to
\be\label{P-odd}
x\e^{x/(2j)} P(\e^{-x/j}) =- x\e^{-x/(2j)} P(\e^{x/j})\,.
\ee
For $x>0$ and $z= \e^{-x/j}$ this relation is equivalent to the reciprocity
condition \re{self}. In the similar manner, if the splitting function verified
the reciprocity relation \re{self}, the corresponding anomalous dimension would
be given at large $N$ by the parity preserving asymptotic series \re{prop}.

Explicit calculation shows~\cite{CFP80,FKL81} that for the nonsinglet twist-two
quark operators the reciprocity relation \re{self} breaks down starting from two
loops and, as a consequence, the asymptotic expansion of the nonsinglet anomalous
dimension involves odd powers of $1/J$ at higher loops. We will demonstrate in
Sect.~3 that the function $f(N)$ related to the nonsinglet anomalous dimension
through \re{f-def} still possesses the parity preserving property \re{prop} to
three loops at least.

\subsection{Singlet anomalous dimensions in QCD}

As the next step, we consider flavour singlet quark and gluon operators of twist
two %. The forward matrix elements of singlet twist-two operators take the form
\ba\label{singlet}
{\mathcal{O}_N^{(q)}(0)} = {\bar q_{\alpha,i}\, \Gamma_{\alpha\beta} D_+^{N-1}
q_{\beta,i}(0)}\,,\qquad {\mathcal{O}_N^{(g)}(0)} = { \tr\left[F_{\mu +}\,
\Pi^{\mu\nu} D_+^{N-2} F_{\nu +}(0)\right]},
\ea
where $F_{\mu+}= F_{\mu+}^a t^a=\ft{i}{g}[D_\mu,D_+]$ is the light-like component
of the gauge field strength tensor and $t^a$ are the generators of the $SU(N_c)$
gauge group in the fundamental (quark) representation. As before, the spinor
matrix $\Gamma_{\alpha\beta}$ and the Lorentz tensor $\Pi_{\mu\nu}$ select `good'
(twist-one) components of the quark and gauge fields. For $\Gamma=\gamma_+$ and
$\Pi_{\mu\nu}=g_{\mu\nu}$ the matrix elements of the operators \re{singlet}
define unpolarized quark and gluon distributions, while for
$\Gamma=\gamma_+\gamma_5$ and $\Pi_{\mu\nu}=\epsilon_{\mu\nu + -}$ they define
longitudinally polarized distributions. % of twist two.

The operators \re{singlet} mix under renormalization with each other and with
operators involving total derivatives. As before, the latter can be eliminated by
going over to the forward matrix elements in \re{singlet}. Then, the matrix
elements $\vev{\mathcal{O}_N^{(q)}(0)}$ and $\vev{\mathcal{O}_N^{(g)}(0)}$ verify
the evolution equation \re{RG} with the mixing matrix $\gamma^{ab}(N)$ (for
$a,b=q,g$) known to two loops for the polarized distributions~\cite{MN95} and to
three loops for the unpolarized distributions~\cite{VMV04}. For our purposes, we
have to examine the eigenvalues of the mixing matrix, $\gamma_q$ and $\gamma_g$.
They can be found as solutions to the characteristic equation
\be\label{ch-eq}
\lr{\gamma_a}^2 - \gamma_a X(N) + Y(N) =0 \,,
%\qquad X = \tr \gamma(N)\,,\qquad Y=\det \gamma(N)
\ee
with the functions $X(N)$ and $Y(N)$ defined as
\ba \nonumber
X(N) &=& \tr \gamma(N) = \gamma^{qq} + \gamma^{gg}\,,
\\[2mm] \label{X,Y}
Y(N) &=& \det \gamma(N) = \gamma^{qq}\gamma^{gg} - \gamma^{qg}\gamma^{gq}\,.
\ea
The explicit expressions for roots to \re{ch-eq} are not particularly suitable
for analyzing the large $N$ asymptotics of the anomalous dimensions  $\gamma_q$
and $\gamma_g$. As we will see in a moment, it is more advantageous to study the
properties of $X(N)$ and $Y(N)$.

As in Sect.~2.1, we first examine the large $N$ expansion of the one-loop mixing
matrix $\gamma^{ab}_0(N)$. Expanding the well-known expression for
$\gamma^{ab}_0(N)$ in powers of the parameter $1/J$ defined in \re{J2}, one finds
%the leading asymptotic behaviour of $\gamma^{ab}_0(N)$ as
\be\label{anom-sing}
\begin{array}{lll}
  \gamma^{qq}_0 = C_F\left[ 2\ln \bar J^2 - 3 + \mathcal{O}(1/J^2)\right],
& & \gamma^{qg}_0 = -2n_f/J + \mathcal{O}(1/J^2), \\[3mm]
 \gamma^{gq}_0 =-2C_F/J + \mathcal{O}(1/J^2),
& & \gamma^{gg}_0 = 2 N_c \ln \bar J^2 - \beta_0 + \mathcal{O}(1/J^2),
\end{array}
\ee
with $\beta_0$ given in \re{beta-QCD}. The matrix elements $\gamma^{qq}_0$ are
the same for the unpolarized and polarized distributions. For $\gamma^{gg}_0$
their difference scales as $\sim 1/J^4$ while for $\gamma^{gq}_0$ and
$\gamma^{qg}_0$ it scales as $\sim 1/J^3$. The large $N$ expansion of the
diagonal matrix elements, $\gamma^{qq}_0$ and $\gamma^{gg}_0$, runs in even
powers of $1/J$ only. Similar expansion of the off-diagonal matrix elements,
$\gamma^{qg}_0$ and $\gamma^{gq}_0$, involves all powers of $1/J$, but odd powers
of $1/J$ disappear in their product $\gamma^{qg}_0\gamma^{gq}_0=4C_Fn_f/J^2 +
\mathcal{O}(1/J^4)$. This implies that the matrix elements $\gamma^{ab}_0(N)$ do
not verify the parity preserving relation \re{prop} but this relation holds true
for the trace and determinant of the mixing matrix, $X(N)$ and $Y(N)$,
respectively. As a consequence, the eigenvalues of the one-loop mixing matrix,
Eq.~\re{ch-eq}, have the same property -- their asymptotic expansion runs
in even powers of $1/J$ only. %
\footnote{Provided that the difference $\gamma^{qq}_0-\gamma^{gg}_0$ scales as
$\mathcal{O}(J^0)$ which is indeed the case for \re{anom-sing} since $C_F\neq
N_c$.} This property can be thought of as a generalization of the relation
\re{gamma0-as} for the singlet anomalous dimensions.

In a close analogy to the nonsinglet case, one can apply \re{f-def} to relate the
two eigenvalues $\gamma_a(N)$ of the mixing matrix to two different functions
$f_a(N)$ (with $a=q,g$).  To the lowest order in the coupling constant, the
functions coincide, $\gamma_a(N)=f_a(N)+\mathcal{O}(\lambda^2)$, and, therefore,
$f_a(N)$ verify the parity preserving property \re{prop} to one loop accuracy. We
will demonstrate in Sect.~3, that for the functions $f_q(N)$ and $f_g(N)$ defined
in this way, the property \re{prop} holds true to three loops at least.

\subsection{Anomalous dimensions in SYM}

Let us now extend the analysis to twist-two operators in supersymmetric
Yang-Mills theories. In the simplest case of the SYM theory with $\mathcal{N}=1$
supercharge, the twist-two operators can be obtained from the singlet operators
\re{singlet} by substituting the quark fields with the gaugino fields belonging
to the adjoint representation of the $SU(N_c)$ group. As a result, the one-loop
anomalous dimensions in the $\mathcal{N}=1$ SYM theory coincide with the QCD
expressions for $C_F=n_f=N_c$~\cite{D77,BFLK85}. In the SYM theories with
$\mathcal{N}>1$ supercharges, construction of twist-two operators becomes
extremely cumbersome due to larger number of fundamental fields involved
(gaugino, gauge fields and scalars). As was shown in Ref.~\cite{BDKM04}, the
problem can be circumvented by employing the formulation of SYM theories in terms
of light-cone superfields~\cite{M83}. This approach takes a full advantage of
superconformal invariance of the SYM theory and provides a unifying description
of the twist-two (superconformal) operators for different number of supercharges
$\mathcal{N}$.

Similar to the QCD case, the form of the one-loop mixing matrix in the SYM theory
is constrained by the (super)conformal invariance of the classical
Lagrangian~\cite{S85}. It allows one to classify the twist-two operators
according to representations of the superconformal $SU(2,2|\mathcal{N})$ group
and, more precisely, its collinear subgroup $SL(2|\mathcal{N})$. In this way, one
can show that, in the SYM theories with $\mathcal{N}<4$ supercharges, {\it all}
twist-two operators belong to {\it two} different $SL(2|\mathcal{N})$ multiplets
whereas in the maximally supersymmetric $\mathcal{N}=4$ SYM theory they belong to
a {\it single} $SL(2|4)$ multiplet. The reason for such difference is that for
$\mathcal{N}=4$ all `good' components of the fundamental fields, providing the
building blocks for the twist-two operators, can be comprised into a single
light-cone superfield whereas for $\mathcal{N}<4$ they are described by two
different light-cone superfields.

The superconformal symmetry ensures that the twist-two operators inside the
supermultiplet have the same anomalous dimension. In the SYM theory with
$\mathcal{N} < 4$ supercharges, the anomalous dimensions of twist-two operators
belonging to two supermultiplets are given to one loop by~\cite{BDKM04}
\ba\label{anom-N}
   \gamma_{0,+}(N) &=& 4\left[\psi(N+1) +\gamma_{\scriptscriptstyle \rm E}\right]-
   \beta_0\,,
   \\ \nonumber
   \gamma_{0,-}(N) &=& 2\left[\psi(N+3-\mathcal{N})+\psi(N-1)
   - (-1)^N \frac{\Gamma(N-1)\Gamma(4-\mathcal{N})}{\Gamma(N+3-\mathcal{N})}+2\gamma_{\scriptscriptstyle \rm E}\right]-
   \beta_0\,,
\ea
with $\beta_0$ being the lowest order coefficient of the beta-function defined in
\re{beta-N}. In the $\mathcal{N}=4$ SYM theory, the twist-two operators belong to
the same supermultiplet and their anomalous dimension is given to one loop
by~\cite{DO02}
\be\label{anom-4}
  \gamma_{0,+}(N) = 4\left[ \psi(N+1) + \gamma_{\scriptscriptstyle \rm
  E}\right].
\ee
Here nonnegative integer $N$ parameterizes the superconformal $SL(2|\mathcal{N})$
spin of the twist-two operators. For the operators with the anomalous dimension
$\gamma_{0,+}(N)$ and $\gamma_{0,-}(N)$ the corresponding quadratic Casimir is
given respectively by~\cite{BDKM04}
\be\label{super-spin}
J_+^2 = N(N+1)\,,\qquad J_-^2 = \lr{N-\ft12 \mathcal{N}}\lr{N+1-\ft12
\mathcal{N}}\,.
\ee
For $\mathcal{N}=0$ the expressions \re{anom-N} coincide with the one-loop
anomalous dimensions of twist-two operators in pure gluodynamics. For
$\mathcal{N}=1$ they match the QCD expressions evaluated in the supersymmetric
limit $C_F=n_f=N_c$~\cite{D77,BFLK85}. In particular, $\gamma_{0,+}(N)$
corresponds to the anomalous dimensions of quark transversity and linearly
polarized gluon distributions~\cite{V97,V98}. For $N=$ even/odd the function
$\gamma_{0,-}(N)$ defines the eigenvalues of the one-loop mixing matrix
$\gamma_0^{ab}(N)$ for singlet unpolarized/polarized distributions~\cite{BFLK85}.

Let us examine the dependence of the anomalous dimensions \re{anom-N} and
\re{anom-4} on the superconformal Casimirs \re{super-spin} at large $N$. The
expression for the anomalous dimension $\gamma_{0,+}(N)$, Eq.~\re{anom-4}, is
similar to \re{anom0} (for $\eta=-1$) and its asymptotic expansion in powers of
$1/J_+^2$ can be easily obtained from \re{gamma0-as}. In the SYM theory with
$\mathcal{N} < 4$ supercharges, the expression for $\gamma_{0,-}(N)$,
Eq.~\re{anom-N}, involves the factor $(-1)^N$ and its analytical continuation
from even and odd $N$ gives rise to two different functions~\cite{CFP80} that we
shall denote as $\gamma_{0}(N,\sigma)$ (with $\sigma=\pm$)
\be\label{gamma-sigma}
\gamma_{0}(N,\sigma)=2 \left[\ln \bar J_-^2+ \varphi_1(J_-^{-2})J_-^{-2} +
\sigma\, \varphi_2(J_-^{-2})\,  J_-^{-4+\mathcal{N}} \right]-\beta_0,
\ee
where, by definition, $\gamma_{0,-}(N)=\gamma_{0}(N,\sigma=(-1)^N)$ for integer
$N$. Here the functions $\varphi_{1,2}$ are given by asymptotic series in
$1/J_-^2$ with the $\mathcal{N}-$dependent coefficients.

We notice that the functions \re{gamma-sigma} have different analytical
properties for even and odd number of supercharges $\mathcal{N}$. For
$\mathcal{N}=0$ and $\mathcal{N}=2$ the functions $\gamma_{0}(N,\sigma)$ admit
asymptotic expansion in integer powers of $1/J_-^2$ only. For $\mathcal{N}=1$
this property is not valid for $\gamma_{0}(N,\sigma)$ anymore but it is restored
in the sum $\gamma_{0}(N,+)+\gamma_{0}(N,-)$. Still, the asymptotic expansion of
the difference $\gamma_{0}(N,+)-\gamma_{0}(N,-)$ runs in odd powers of $1/J_-$
only.

To understand the reason why the parity preserving property is modified for
$\mathcal{N}=1$, we recall that $\gamma_{0}(N,\sigma)$ defines the eigenvalues of
the singlet mixing matrix in QCD, Eq.~\re{anom-sing}, in the supersymmetric limit
$C_F=n_f=N_c$. More precisely, for integer $N$ the solutions to the
characteristic equation \re{ch-eq} are given by $\gamma_1=\gamma_{0,-}(N+1)$ and
$\gamma_2=\gamma_{0,-}(N)$. Then, it follows from \re{ch-eq} and \re{X,Y} that
\be\label{sqrt}
\gamma_1-\gamma_2=
\sqrt{\lr{\gamma_0^{qq}-\gamma_0^{gg}}^2+4\gamma_0^{qg}\gamma_0^{gq}}
\ee
with the matrix elements of the mixing matrix given by \re{anom-sing}. We
demonstrated in Sect.~2.2, that $\lr{\gamma_0^{qq}-\gamma_0^{gg}}^2$ and
$\gamma_0^{qg}\gamma_0^{gq}$ are given by asymptotic series in $1/J^2=1/(N(N+1))$
and, therefore, $\gamma_1-\gamma_2$ should also admit similar expansion provided
that the leading $\sim J^0$ term inside the square-root in the right-hand side of
\re{sqrt} is different from zero, or equivalently $\gamma_1-\gamma_2\neq 0$ at
large $N$. In the $\mathcal{N}=1$ SYM theory, the eigenvalues of the mixing
matrix become degenerate in the large $N$ limit and the above condition is not
satisfied. Indeed, one finds from \re{anom-sing} that
$\lr{\gamma_0^{qq}-\gamma_0^{gg}}^2+4\gamma_0^{qg}\gamma_0^{gq}\sim 1/J^2$ in the
supersymmetric limit, $C_F=n_f=N_c$, leading to \footnote{The phenomenon is
well-known in quantum mechanics for the system of two almost degenerate energy
levels $\gamma_0^{gg}$ and $\gamma_0^{qq}$ whose interaction energy is of the
same order as the level splitting, $\gamma_0^{qg}\gamma_0^{gq}\sim
\lr{\gamma_0^{qq}-\gamma_0^{gg}}^2\sim 1/J^2$.}
\be\label{diff-gam}
\gamma_{0}(N,\sigma)-\gamma_{0}(N+1,-\sigma) = J^{-1} \varphi(J^{-2})
\ee
with $J^2=N(N+1)$ and $\varphi(J^{-2})$ given by series in $1/J^2$. For
$\mathcal{N=}1$, one replaces $\gamma_{0}(N,\sigma)$ by its general expression
\re{gamma-sigma} (without assuming parity property of the function $\varphi_2$),
takes into account the relation $J_-^2=(N-\ft12)(N+\ft12)$, Eq.~\re{super-spin},
and deduces from \re{diff-gam} that, in agreement with \re{gamma-sigma}, the
$\sigma-$dependent term inside $\gamma_{0}(N,\sigma)$ should admit an asymptotic
expansion in odd powers of $1/J_-$ only.

\section{Parity preserving relations at higher loops}

We demonstrated in Sect.~2, that to one-loop accuracy, the large $N$ expansion of
the twist-two anomalous dimensions in QCD and in its supersymmetric extensions
(except the $\mathcal{N}=1$ SYM theory) runs in integer powers of the
(super)conformal Casimir $J^2$ and, therefore, possesses the parity preserving
property \re{prop}. In the $z-$representation \re{z-repr}, this translates into
the reciprocity relation \re{self} for the corresponding one-loop splitting
functions $P(z)$. Going over to higher loops, one finds that the conformal spin
of Wilson operators get renormalized by an amount proportional to their anomalous
dimension \re{J-ren}. Moreover, in gauge theories with nonvanishing beta-function
the anomalous dimensions receive conformal symmetry breaking contributions. We
will argue in this section, that once both effects are taken into account, the
parity preserving relation \re{self} does not hold true for the anomalous
dimension $\gamma(N)$ but it remains valid (to three loops at least) for the
function $f(N)$ defined in \re{f-def}.

\subsection{Space-like anomalous dimensions}

Let us examine the relation between the space-like anomalous dimension of twist
$L$ and the function $f(N)$
\be\label{L-def}
\gamma_S(N) = f\left(N+\ft12\gamma_S(N) -\ft14 L\beta(\lambda)\right)\,.
\ee
We recall that the shift of the argument in the right-hand side of \re{L-def}
takes into account renormalization of the conformal spin of the quasipartonic
operator of twist $L$ due to nonzero value of the anomalous dimension and of the
beta-function. The function $f(N)$ admits a perturbative expansion in the
coupling constant $\lambda$ similar to \re{gamma-exp}
\be\label{f-exp}
f(N) = \lambda f_0(N) + \lambda^2 f_1(N) + \ldots\,.
\ee
To the lowest order one has $f_0(N)=\gamma_0(N)$.
% due to vanishing of $\gamma_S(N)$ and $\beta(\lambda)$ for $\lambda\to 0$.
Substituting \re{f-exp} into \re{L-def} and matching the coefficients in front of
powers of $\lambda$, one can establish the relations between higher order
corrections to $f(N)$ and $\gamma_S(N)$. It is convenient to introduce two
functions
\ba
\nonumber %\label{f-tilde}
\widehat\gamma_S(N) &=& \gamma_S(N)- \ft12 L{\beta(\lambda)}\,,
\qquad \\[2mm] \label{f-tilde}
\widehat f(N) &=& f(N)- \ft12 L{\beta(\lambda)}\,,
\ea
so that the relation \re{L-def} simplifies as
\be\label{tilde-gamma}
\widehat\gamma_S(N) = \widehat f\left(N +\ft12\widehat\gamma_S(N)\right)\,.
\ee
%where the function $\widehat f(N)$ still depends on $\beta(\lambda)$.
Solving this functional relation one obtains its formal solution as
(Lagrange-B\"urmann formula)
\be \label{gamma-series} %\nonumber
\widehat\gamma_S(N) = \sum_{k=1}^\infty \frac{1}{k!}\left(\ft12\partial_N
\right)^{k-1} [\widehat f\,(N)]{}^k = \widehat f(N) +\frac14\lr{\widehat
f{}^2(N)}'+\frac1{24} \lr{\widehat f{}^3(N)}'' + \mathcal{O}(\widehat f^4(N))\,,
\ee
where prime denotes a derivative with respect to $N$. To determine the function
$\widehat f(N)$ one has to invert this relation. To this end, one uses
\re{tilde-gamma} to get
\be\label{f-hat0}
\widehat f(N)=\widehat\gamma_S(N-\ft12\widehat f(N))
\ee
and obtains solution to this equation as
\be \label{f-hat}
\widehat f(N) =
\sum_{k=1}^\infty \frac{1}{k!}\left(-\ft12\partial_N
%\frac{d }{d N}
\right)^{k-1} [\widehat \gamma_S(N)]{}^k = \widehat \gamma_S(N)
-\frac14\lr{\widehat \gamma{}_S^2(N)}'+\frac1{24} \lr{\widehat \gamma{}_S^3(N)}''
+ \mathcal{O}(\widehat \gamma_S^4(N))\,.
\ee
This relation allows one to reconstruct the function $\widehat f(N)$ from known
expression for the space-like anomalous dimensions $\widehat \gamma_S(N)$.

\subsubsection{Parity preserving relations}

We are now ready to verify the parity preserving relation \re{prop} for the
scaling function $f(N)$ in higher loops. For the {\it nonsinglet} anomalous
dimensions, the analysis goes along the same lines as in Sect.~2.1 -- one
replaces $\gamma_S(N)$ in \re{f-hat} and \re{f-tilde} by available two- and
three-loop nonsinglet anomalous dimensions, evaluates the function $f(N)$ to the
same loop accuracy and, then, examines its asymptotic expansion in $J^2=N(N+1)$
at large $N$. For the {\it singlet} anomalous dimensions, one uses two solutions
to the characteristic equation \re{ch-eq}, $\gamma_q(N)$ and $\gamma_g(N)$, and
defines the corresponding functions $f_q(N)$ and $f_g(N)$ with a help of
\re{f-hat} (see Appendix for more details).

In our analysis, we used expressions for multi-loop anomalous dimensions of
various twist-two operators in QCD and  in SYM theories available in the
literature. They include
\begin{itemize}
  \item Two-loop longitudinally polarized singlet distributions in QCD~\cite{MN95};
  \item Two-loop gluon linearly polarized distribution in QCD~\cite{V98};
  \item Two-loop quark transversity distribution in QCD~\cite{V97} and its analogs
in SYM theories with $\mathcal{N}=0,1,2$ supercharges~\cite{BKM04};
  \item Three-loop nonsinglet unpolarized distributions in QCD~\cite{MVV04};
  \item Three-loop singlet unpolarized distributions in QCD~\cite{VMV04};
  \item Three-loop `universal' distribution in $\mathcal{N}=4$ SYM~\cite{KLOV04}.
\end{itemize}
These anomalous dimensions are given by complicated expressions involving nested
harmonic sums and various color factors. Going over through a lengthy
calculation, we worked out their large $N$ expansion in powers of
$J^{-1}=\lr{N(N+1)}^{-1/2}$ and split the corresponding series as
\be\label{sum-gamma}
\gamma_S(N) = \gamma_+ (\ln \bar J, J^{-1}) +  \gamma_- (\ln
\bar J, J^{-1})+\ft12 L \beta(\lambda)\,.
\ee
Here the beta-function term is introduced for the later convenience and
$\gamma_\pm$ are series in $J^{-1}$ of a definite parity, $\gamma_\pm (\ln \bar
J, -J^{-1})=\pm \gamma_\pm (\ln \bar J,J^{-1})$,
\ba \nonumber
\gamma_+ &=& \Gamma_0(\ln \bar J) + \Gamma_2(\ln\bar J)\,\b{J^{-2}} + % \Gamma_4(\ln J)\, J^{-4} +
\mathcal{O}(\b{J^{-4}})\,,
\\[2mm] \label{Gamma_k}
\gamma_- &=& \Gamma_1(\ln \bar J)\, \b{J^{-1}} + \Gamma_3(\ln \bar J)\,
\b{J^{-3}} + \mathcal{O}(\b{J^{-5}})\,,
\ea
where $\Gamma_k(\ln \bar J)$ are given by series in the coupling constant with
the coefficients depending on $\ln \bar J=\ln J +\gamma_{\rm\scriptscriptstyle
E}$. Matching \re{sum-gamma} and \re{DMS-as} one can establish the correspondence
between the two sets of coefficient functions (for $L=2$)
\ba\nonumber
\Gamma_0 &=& A(\lambda)\ln \bar J+B(\lambda)-\beta(\lambda)\,,\qquad
\\[2mm] \label{Gamma-A}
\Gamma_1 &=& C_-(\lambda)\ln\bar J + D_-(\lambda) \,.
\ea
In all cases except the $\mathcal{N}=4$ SYM, the resulting multi-loop expressions
for the coefficient functions $\Gamma_k(\ln\bar J)$ are too cumbersome to be
displayed here. For the three-loop `universal' anomalous dimension our analysis
gives (see Appendix for details)
\ba \nonumber
\Gamma_0&=&4\,\alpha\,\ln\bar J -6\,{\alpha}^{2}\,\zeta_3  + {\alpha}^{3}\left(
-\ft43\,{\pi }^{2}\zeta_3 +20\,\zeta _5 \right)+\mathcal{O}(\alpha^4)\,,
\\[2mm] \nonumber
\Gamma_1&=&  {8\,{\alpha}^{2}\,\ln\bar J
-12\,{\alpha}^{3}}\,\zeta_3+\mathcal{O}(\alpha^4)\,,
\\[1mm] \nonumber
\Gamma_2&=&{\ft23\,\alpha+2\,{\alpha}^{2}+ {\alpha}^{3}}\left( \ft23\,{\pi }^{2}
+\lr{16-\ft23\,{\pi }^{2}}\,\ln\bar J -8\, \left( \ln\bar J \right)
^{2}\right)+\mathcal{O}(\alpha^4)\,,
\\[1mm] \label{f-N=4}
\Gamma_3&=&{ {\alpha}^{2}\left(\ft43-\ft53\,\ln\bar J  \right) +
{\alpha}^{3}}\left( 4+\ft52\,\zeta_3 -8\,\ln\bar J
\right)+\mathcal{O}(\alpha^4)\,, \quad \ldots
\ea
where $\alpha$ stands for the so-called physical coupling constant~\cite{CMW91}
\be\label{alpha}
\alpha=\ft12\Gamma_{\rm cusp}(\lambda)=\lambda-\ft16\,{\pi
}^{2}{\lambda}^{2}+{\ft {11}{180}}\,{\pi }^{4}{ \lambda}^{3} +
\mathcal{O}(\lambda^4)\,.
\ee
Replacing $\gamma_S(N)$ in the relation \re{f-hat} by its expression,
Eqs.~\re{sum-gamma}, \re{Gamma_k} and \re{f-N=4}, one finds the scaling function
in the $\mathcal{N}=4$ SYM theory after some algebra as
\ba\nonumber
f^{(\mathcal{N}=4)}(N)&=&4\,\alpha\ln\bar J  -6\, {\alpha}^{2} \zeta_3 +
{\alpha}^{3}\left( -\ft43\,{\pi }^{2}\zeta_3 +20\,\zeta_5  \right)
\\[2mm] \label{OK}
&+& \left[\ft23\,\alpha+2\,\alpha^{2}+
 \left( \ft23\,{\pi }^{2}-\ft23\,{\pi }^{2}\ln\bar J
 \right) {\alpha}^{3}\right]\b{{J}^{-2}}+\mathcal{O} \left( \b{{J}^{-4}},\alpha^4
 \right)\,.
\ea
We observe that, in agreement with \re{prop}, this series does not involve odd
powers of $1/J$ while they are present in the series for the anomalous dimension
\re{sum-gamma}.

We repeated similar analysis for the above-mentioned two-loop and three-loop
anomalous dimensions of twist two and found that in all cases the corresponding
functions $f(N)$ verify the parity preserving relation \re{prop}! The properties
of the functions $f(N)$ are discussed in the Appendix.

\subsubsection{MVV relations}

As was mentioned in the Introduction, the parity preserving property \re{prop}
implies certain relations between different terms in the large $N$ expansion of
the anomalous dimensions. To obtain these relations, one notices that according
to \re{prop} the large $N$ expansion of the derivatives $\lr{\partial_N}^n f =
(\sqrt{1+1/{(4J^2)}}\partial_J)^n f$ runs in odd/even powers of $1/J$ depending
on parity of $n$. Matching \re{gamma-series} and \re{sum-gamma} one finds that
the series  $\gamma_+$ and $\gamma_-$ are given by the sum of terms involving
derivatives of the even and odd order, respectively,
\ba \nonumber
\gamma_+ &=& \sum_{k=0}^\infty \frac{1}{(2k+1)!}\left(\ft12\partial_N
\right)^{2k} [\widehat f(N)]{}^{2k+1}=\widehat f(N) + \mathcal{O}(\lambda^3)\,,
\\ \label{gamma+-}
\gamma_- &=&\sum_{k=1}^\infty \frac{1}{(2k)!}\left(\ft12\partial_N \right)^{2k-1}
[\widehat f(N)]{}^{2k}=\ft14 \partial_N[\widehat f(N)]^2 +
\mathcal{O}(\lambda^4)\,.
\ea
Inverting the first relation and substituting the resulting expression for $f(N)$
into the second relation, one can express $\gamma_-$ in terms of $\gamma_+$. To
four-loop accuracy one finds
\be\label{Gamma-rel}
\gamma_- = \ft14 \lr{{\gamma}_+^{\,2}}'+\ft1{48}\left(-{\gamma}_+
\lr{{\gamma}_+^3}'' +\ft1{4} \lr{{\gamma}_+^4}''\right)' +\ldots
\ee
where prime denotes a derivative with respect to $N$ and ellipses stand for terms
involving higher number of $\gamma_+$ and their derivatives with respect to $N$.
It follows from \re{Gamma-rel} that in order to determine $\gamma_-$ to four
loops it is sufficient to substitute $\gamma_+$ by its three-loop expression.

Replacing $\gamma_\pm$ in \re{Gamma-rel} by their expressions \re{Gamma_k} and
comparing the coefficients in front of odd powers of $1/J$ one obtains an
infinite system of relations between the coefficient functions $\Gamma_k$
entering \re{Gamma_k}
\ba\nonumber
\Gamma_1 &=& \ft12  \Gamma_0\,\dot{\Gamma}_0\,,
\\[2mm] \label{G1}
\Gamma_3 &=& -\ft1{12}\Gamma_0^3\, \dot{\Gamma}_0+\ft14\Gamma_0^2
\big(\dot{\Gamma}_0\big)^2-\ft18 \Gamma_0\big(\dot{\Gamma}_0\big)^3
+\ft1{16}{\Gamma}_0\dot{\Gamma}_0+\ft12 \dot{\big(\Gamma_0\Gamma_2\big)} -
\Gamma_0\Gamma_2\,, \quad \dots\,,
\ea
where dot denotes a derivative with respect to $\ln\bar J$. Here, in the second
relation, we took into account that $\Gamma_0$, Eq.~\re{Gamma-A}, is linear in
$\ln\bar J$ and therefore $\ddot{\Gamma}_0=0$. It is straightforward to derive
from \re{Gamma-rel} similar relations for the subleading functions $\Gamma_5,
\Gamma_7, \ldots$ but they become more cumbersome. Substitution of \re{Gamma-A}
into the first relation in \re{G1} yields $C_-=\ft12 A^2$ and $D_-=\ft12
A(B-\beta)$, in agreement with the exact three-loop result in QCD,
Eqs.~\re{DMS-ACD} and \re{D-exact}. In the $\mathcal{N}=4$ SYM theory, the
relations \re{G1} can be easily verified with a help of \re{f-N=4}.

We conclude from \re{Gamma-rel} and \re{G1} that the coefficients in front of odd
powers of $1/J$ in the large $N$ expansion of the anomalous dimension,
Eqs.~\re{sum-gamma} and \re{Gamma_k}, can be expressed in terms of the
coefficients accompanying smaller {\it even} powers of $1/J$ to {\it less} number
of loops.

\subsection{Time-like anomalous dimensions}

Let us now extend consideration to the time-like anomalous dimensions
$\gamma_T(N)$. They describe the scale dependence of the partonic fragmentation
functions~\cite{CTEQ} and, in distinction with the space-like anomalous
dimensions discussed above, are not related to local Wilson operators. In what
follows we shall assume, following~\cite{DMS}, that the time-like anomalous
dimensions satisfy the relation
\be\label{gamma-T}
\gamma_T(N) = f\left(N-\ft12\gamma_T(N) -\ft14 L\beta(\lambda)\right)\,.
\ee
with the function $f(N)$ {\it the same} as for the space-like anomalous dimension
\re{L-def}. The relation \re{gamma-T} generalizes to higher twists the relation
\re{DMS} proposed in \cite{DKT95,DMS} and incorporates the beta-function
contribution. Inverting \re{gamma-T} and combining it with the similar relation
for the space-like anomalous dimension \re{L-def}, one obtains
\ba\label{f-aux}
f(N)= \gamma_T\left(N + \ft12f(N) +\ft14 L{\beta(\lambda)}\right)=
\gamma_S\left(N
 -\ft12f(N)+\ft14 L{\beta(\lambda)}\right)\,.
\ea
One can exclude $f(N)$ from \re{f-aux} and obtain two equivalent functional
relations between the space-like and time-like anomalous dimensions
\ba \nonumber
\gamma_S(N) &=& \gamma_T(N+\gamma_S(N))\,,
\\[2mm] \label{gamma-T-S}
\gamma_T(N) &=& \gamma_S(N-\gamma_T(N))\,.
\ea
Notice that the beta-function term dropped out from these relations. The second
relation in \re{gamma-T-S} takes the same form as \re{M} but in distinction with
the latter it now holds true for arbitrary $N$.

Following \cite{DMS}, one can apply the relations \re{gamma-T-S} to verify the
reciprocity relations \re{GL}. It is convenient to switch in \re{GL} from the
splitting functions to their moments \re{z-repr} and examine the relations
between the anomalous dimensions $\gamma_S(N)$ and $\gamma_T(N)$. From the second
relation in \re{gamma-T-S} one gets similarly to \re{f-hat0} and \re{f-hat}
\be
\gamma_T(N) = \sum_{k=1}^\infty \frac{1}{k!}\left(-\partial_{N }\right)^{k-1}
[\gamma_S(N)]^k\,.
\ee
We conclude from this relation that the difference between the space-like and
time-like anomalous dimensions is given by
\be\label{diff-gamma}
\gamma_T(N) - \gamma_S(N) = -\frac12\lr{\gamma_S^2(N)}'+\frac1{6} \lr{
\gamma_S^3(N)}'' -\frac1{24} \lr{ \gamma_S^4(N)}'''
+\mathcal{O}(\gamma_S^5(N))\,,
\ee
where prime denotes a derivative with respect to $N$. It is different from zero
starting from two loops. Moreover, replacing $\gamma_S(N)$ in the right-hand side
of \re{diff-gamma} by its three-loop expression, one can evaluate $\gamma_T(N) -
\gamma_S(N)$ to four-loop accuracy and, then, translate it with a help of
\re{z-repr} into the difference of the splitting functions $P_S(z)-P_T(z)$.

Let us now examine the second reciprocity relation in \re{GL} and evaluate the
moments \re{z-repr} of the difference of the splitting functions entering both
sides of \re{GL}
\be\label{Delta0}
\Delta(N)= -\int_0^1 dz\, z^{N-1} \left[P_T(z)+z P_S(1/z) \right]\,.
\ee
By definition \re{z-repr}, the moments of $P_T(z)$ give rise to the time-like
anomalous dimension $\gamma_T(N)$ which can be expressed in its turn in terms of
the function $f(N)$ using \re{gamma-T}. %
\footnote{One observes that the relation \re{gamma-T} transforms into \re{L-def}
under substitution $\gamma_S(N) \to -\gamma_T(N)$ and $f(N) \to -f(N)$. The
expression for $\gamma_T(N)$ can be easily obtained by applying the same
transformation to \re{gamma-series} and \re{f-tilde}.} To calculate the moments
of $P_S(1/z)$ in \re{Delta0}, we go over to the large $N$ limit and change the
integration variable $z= \e^{-x/j}$ with $j=N+\ft12=\lr{J^2+\ft14}^{1/2}$ as in
\re{sample}. Expanding the $x-$integral in inverse powers of $J$, one finds that
it is given by the asymptotic series \re{sum-gamma} with $J^{-1}\to -J^{-1}$
\be
\int_0^1 dz\, z^{N} P_S(1/z) =\gamma_+ (\ln \bar J, J^{-1}) -  \gamma_- (\ln
\bar J, J^{-1})+\ft12 L \beta(\lambda)\,.
\ee
Then, one takes into account \re{gamma+-} and obtains
\be\label{Delta}
\Delta(N) = \sum_{k=2}^\infty \frac1{k!} \lr{-\ft12 \partial_N}^{k-1} \left\{
\left[f(N)+\ft12  L{\beta(\lambda)}\right]^k-\left[f(N)- \ft12
L{\beta(\lambda)}\right]^k \right\}\,.
\ee
It follows from this relation that $\Delta(N)$ vanishes in a conformal gauge
theory with $\beta(\lambda)=0$. Otherwise, $\Delta(N)$ receives corrections
starting from two loops.

The relation \re{Delta} takes a simple form when expressed in terms of the
time-like anomalous dimension $\gamma_T(N)$. Inverting \re{gamma-T} and
substituting $f(N)$ in \re{Delta} by its expression in terms of $\gamma_T(N)$,
one gets after some algebra
\be
\Delta(N) =\gamma_T(N)-\gamma_T(N+\ft12 L\beta(\lambda))\,.
\ee
Its expansion in powers of the beta function reads
\be
\Delta(N)=- \ft12 L\beta(\lambda)\left[\gamma_T'(N)+{\ft
{1}{4}}{{L}{\beta}(\lambda)} \gamma_T''(N)  +{\ft
{1}{24}}\lr{{L}{\beta}(\lambda)}^2 \gamma_T'''(N)\right] +
\mathcal{O}(\lambda^5)\,.
\ee
We observe that this relation has the same remarkable property as
Eqs.~\re{Gamma-rel} and \re{diff-gamma} -- using the known result for three-loop
anomalous dimensions, one can determine $\Delta(N)$ to four loops.

Our analysis in this section relied on the parity preserving property \re{prop}
of the scaling function $f(N)$. We verified this property in QCD and in SYM
theories using known results for the multi-loop anomalous dimensions. In the next
two sections, we will argue that the property \re{prop} holds true to all loops
in QCD in the so-called large $\beta_0$ limit and in the $\mathcal{N}=4$ SYM
theory for special class scalar operators of higher twist in the strong coupling
regime.

\section{Parity preserving relations in the large $\beta_0-$limit}

The anomalous dimensions in QCD depend on the number of quark flavours $n_f$. One
can make use of this fact to obtain the all-loop expression for the anomalous
dimensions $\gamma(N)$ and the functions $f(N)$ in the large $\beta_0$ limit
\be\label{a}
a= \ft12 \beta_0 \lambda =\mbox{fixed} \,,\qquad \beta_0 \to\infty\,.
\ee
Here $\beta_0=\ft{11}3 N_c-\ft23 n_f$ is the one-loop coefficient of the
beta-function in QCD and $n_f$ is assumed to take large negative values. Notice
that this limit does not exist in SYM theories, since $\beta_0$ is uniquely fixed
there by the number of supercharges, Eq.~\re{beta-N}.

Let us first consider the nonsinglet (unpolarized) twist-two anomalous dimension
in QCD. To one-loop order, it takes the form $\gamma_{\rm ns} = \lambda
\gamma_0(N)+\mathcal{O}(\lambda^2)$ with $\gamma_0(N)$ given by \re{anom0} for
$\eta=1$. At higher loops, it receives corrections $\sim \lambda (\lambda n_f)^n$
enhanced in the large $\beta_0$ limit. They come from the so-called single
renormalon chain Feynman diagrams which can be resummed to all orders in the
rescaled coupling constant $a$, Eq.~\re{a}. Then, the nonsinglet anomalous
dimension in QCD is
given in the large $\beta_0-$limit by~\cite{Gr96,G05} % (see Eq.~(C.1) in hep-ph/0501257)
\be\label{gamma-inf}
\gamma_{\rm ns}^{(\infty)}(N)=2 A(a)\left[\psi \left( N+a \right) -\psi \left(
1+a \right) +\frac{N-1}2 \left( {\frac {{a}^{2}+2\,a-1}{ \left( 1+a \right)
\left( N +a \right) }}-{\frac { \left( 1+a \right) ^{2}}{ \left( a+2 \right)
 \left( N+1+a \right) }} \right)\right]
\ee
where the superscript $(\infty)$ indicates the leading asymptotic behaviour for
$\beta_0\to\infty$ and $A(a)$ is the cusp anomalous dimension in the limit \re{a}
\be
A(a) = {\frac {2{\it C_F}\,\sin \left( \pi \,a \right) \Gamma \left( 4+2 \,a
\right) }{3\beta_{{0}}\pi \, \Gamma^{2} \left( a+2 \right)}}\,.
\ee
To lowest order in $a$,  $\gamma^{(\infty)}_{\rm ns}(N)$ coincides with the exact
one-loop result \re{anom0} (for $\eta=1$).

Let us substitute \re{gamma-inf} into the relation \re{L-def} and reconstruct the
corresponding function $f^{(\infty)}(N)$. One finds that the relation  \re{L-def}
simplifies significantly in the large$-\beta_0$ limit because the anomalous
dimension scales as $\gamma_{\rm ns}^{(\infty)}(N) \sim 1/\beta_0$ and,
therefore, it can be safely neglected in the right-hand side of \re{L-def}.
Together with $\beta(\lambda) = -\beta_0\lambda+\ldots = - 2a + \ldots$ this
leads to
\be\label{gamma-infinity}
\gamma^{(\infty)}_{\rm ns}(N) = f_{\rm ns}^{(\infty)}\left(N +a \right).
\ee
As a result, one finds from \re{gamma-inf} after some algebra
\be\label{f-infinity}
f_{\rm ns}^{(\infty)}(N)=2 A(a)\bigg[ \psi \left(N+1\right) -\psi \left( 1+a
\right) -
 {\frac {\left(2a+1+\eta \right) ^{2}}{8N(N+1) }}
- {\frac {3-{a}^{2}}{2 \left( 1+a \right)  \left(2+a
 \right) }} \bigg]\,,
\ee
with $\eta=1$. The reason why we displayed the $\eta-$dependence in
\re{f-infinity} is that for $\eta=-1$ the relations \re{f-infinity} and
\re{gamma-infinity} describe the quark transversity anomalous dimension,
Eq.~\re{anom0}, in the large $\beta_0$ limit~\cite{Gr03}. As expected, the
function $f^{(\infty)}(N)$ also depends on the parameter $\varepsilon_{\rm
cr}=\beta(\lambda)/2=-a+\ldots$ defining the `critical' value of the space-time
dimension $d_{\rm cr}=4-2\varepsilon_{\rm cr}$ for which the gauge theory becomes
conformal. Since $\gamma_{\rm ns}^{(\infty)}(N)\sim 1/\beta_0$, one deduces from
\re{gamma-T-S} that the space-like and time-like anomalous dimensions coincide in
the large $\beta_0$ limit and, therefore, the first of the two reciprocity
relations in \re{GL} is exact.

We recall that the expression \re{f-infinity} resums perturbative corrections to
the function $f_{\rm ns}(N)$ of the form $\sim \lambda \lr{\lambda n_f}^n$. The
$N-$dependence of the function $f_{\rm ns}^{(\infty)}(N)$, Eq.~\re{f-infinity},
is similar to that of the one-loop expression \re{anom0} -- higher order
corrections in $a$ only modify the coefficients in front of $\psi(N+1)$ and
$1/(N(N+1))$. Therefore, in agreement with \re{prop}, the large $N$ expansion of
$f_{\rm ns}^{(\infty)}(N)$ runs in integer negative powers of $J^2=N(N+1)$ only.

Let us now examine the large $\beta_0$ limit of the twist-two (un)polarized
singlet anomalous dimensions, $\gamma_q$ and $\gamma_g$, defined as solutions to
the characteristic equation \re{ch-eq}. To one-loop accuracy, the mixing matrix
is given by \re{anom-sing} and the $n_f-$dependence resides in two matrix
elements only, $\gamma^{qg}_0$ and $\gamma^{gg}_0$. Then, it is easy to see that
eigenvalues of the one-loop mixing matrix scale in the large $\beta_0$ limit
\re{a} as
\be\label{finite}
\gamma_q^{(\infty)}(N) \sim  \lambda = a/\beta_0  \,,\qquad
\gamma_g^{(\infty)}(N) = -\lambda \beta_0 = -2a\,.
\ee
Going over to higher loops one finds~\cite{Gr96,BG97} that leading corrections to
both anomalous dimensions take the form $\sim \lambda (\lambda n_f)^n\sim
a^{n+1}/\beta_0$. The scaling behaviour of the smallest eigenvalue
$\gamma_q^{(\infty)}(N)$ is the same as of the nonsinglet anomalous dimension
$\gamma_{\rm ns}^{(\infty)}(N)$. Therefore, in the large $\beta_0$ limit the
relation \re{L-def} between $\gamma_q(N)$ and the corresponding function $f_q(N)$
takes the same form as \re{gamma-infinity}
\be\label{f_q}
\gamma^{(\infty)}_{q}(N) = f^{(\infty)}_q\left(N +a \right).
\ee
In distinction with $\gamma^{(\infty)}_{q}(N)$, the gluon dominated anomalous
dimension $\gamma_g^{(\infty)}(N)$ approaches a finite value \re{finite} and the
defining relation \re{L-def} looks in this case as
\be
\gamma^{(\infty)}_{g}(N) = f^{(\infty)}_g\left(N +\ft12\gamma^{(\infty)}_{g}(N)
+a \right)\,.
\ee
Substituting $\gamma_g^{(\infty)}(N)=-2a + \mathcal{O}(1/\beta_0)$ into this
relation, one finds in the large $\beta_0$ limit
\be\label{f_g}
\gamma^{(\infty)}_{g}(N) = f^{(\infty)}_g(N)\,,
\ee
with $f^{(\infty)}_g(N)+2a = \mathcal{O}(1/\beta_0)$. The relations \re{f_q} and
\re{f_g} hold true for the unpolarized and polarized anomalous dimensions
although the corresponding scaling functions $f_q(N)$ and $f_g(N)$ are different
in two cases.

The resummed  (un)polarized singlet anomalous dimensions
$\gamma^{(\infty)}_{q}(N)$ and $\gamma^{(\infty)}_{g}(N)$ have been calculated in
Refs.~\cite{Gr96,BG97}. \footnote{Notice that the expression for the quark
dominated polarized anomalous dimensions, Eq.(5.3) in the second reference in
\cite{Gr96}, contains a misprint. The factor $(2\mu+n-1)$ should be replaced
there by $(2\mu+n-2)$. We would like to thank J.~Gracey for clarifying this
point.} Their expressions are lengthy and to save space we do not display them
here. Making use of \re{f_q} and \re{f_g}, it is straightforward to calculate the
corresponding functions $f^{(\infty)}_q(N)$ and $f^{(\infty)}_g(N)$ and, then,
work out their large $N$ expansion. In this way, we found that for the polarized
and unpolarized anomalous dimensions the expansion of $f^{(\infty)}_q(N)$ and
$f^{(\infty)}_g(N)$ runs in integer negative powers of $J^2=N(N+1)$ only! We
conclude that the parity preserving property \re{prop} holds true in QCD for the
twist-two (non)singlet anomalous dimensions to all loops in the large $\beta_0$
limit.

\section{Parity preserving relations in the AdS/CFT}

In this section, we will employ the AdS/CFT correspondence to determine the
function $f(N)$ in the maximally supersymmetric $\mathcal{N}=4$ theory in the
strong coupling regime.

The gauge/string duality allows one to establish the correspondence between the
anomalous dimensions of Wilson operators in the planar $\mathcal{N}=4$ SYM theory
at strong coupling and energies of strings on the AdS${}_5\times$S${}^5$
background~\cite{Mal97,BMN02,GKP02}. This relationship takes a very precise form
for Wilson operators built from a complex scalar field $Z$ and the light-cone
component of the covariant derivatives $D_+ = (n\cdot D)$ (see review \cite{T03}
and references therein)
\be\label{O}
\mathcal{O}^{\{k\}}_{L}(0) = \tr \left[D_+^{k_1} Z(0) D_+^{k_2} Z(0) \ldots
D_+^{k_L} Z(0) \right],
\ee
with $k_i$ being nonnegative integer. In gauge theory,
$\mathcal{O}^{\{k\}}_{L}(0)$ can be identified as quasipartonic operators with
twist $L$ and the Lorentz spin $N=\sum_i k_i$. The operators \re{O} with the same
quantum numbers $L$ and $N$ mix under renormalization and their anomalous
dimensions can be found by diagonalizing the corresponding mixing matrix. For
$L=2$, the anomalous dimension of the operators \re{O} is uniquely specified by
the total Lorentz spin $N$ whereas for higher twists $L\ge 3$ the anomalous
dimensions occupy a band whose size depends on $N$ (see review \cite{BBGK04} and
references therein). The AdS/CFT correspondence provides a prediction for the
{\it minimal} anomalous dimension $\gamma_L(N)$ of the Wilson operators \re{O}
carrying large quantum numbers (twist $L$ and/or Lorentz spin $N$) in the planar
$\mathcal{N}=4$ SYM theory at strong coupling. Namely, $\gamma_L(N)$ is related
to the energy of a (semi)classical folded string which rotates with the angular
momentum $N$ on the AdS${}_3$ part of the target space and whose center-of-mass
is also moving with the angular momentum $L$ along a big circle of
S${}^5$~\cite{GKP02,FT03}
\be\label{ads}
\gamma_L(N)=E-N-L\,.
\ee
Here the string energy $E$ is given by the series in $1/\sqrt{\lambda}$
\be\label{E0}
E = \sqrt{\lambda} E_0\lr{\frac{N}{\sqrt{\lambda}},\frac{L}{\sqrt{\lambda}}} +
 E_1\lr{\frac{N}{\sqrt{\lambda}},\frac{L}{\sqrt{\lambda}}}+\ldots
\ee
with $E_0$ being the classical string energy and $E_1$ being one-loop quantum
string correction.

For the folded rotating string, the classical energy $E_0$ can be found
explicitly in terms of the Jacobi elliptic $\mathbb{K}-$ and
$\mathbb{E}-$functions~\cite{FT03,BFST03,KZ04}. The quantum correction $E_1$ has
been recently calculated in Ref.~\cite{FTT06} and its expression is more
involved. Keeping in \re{E0} the first term only, one finds from \re{ads} the
leading order expression for the anomalous dimension as $\lambda \gg 1$ and
${N}/{\sqrt{\lambda}},\ {L}/{\sqrt{\lambda}}={\rm fixed}$ \cite{BGK06,SS06}
\ba\label{N/L}
&& \gamma_L(N) =  L\left[ \mathbb{K}(\tau)\, b_{\rm str} - \mathbb{E}(\tau)
\frac{{\lambda'}}{b_{\rm str}} - 1\right] \, , \nonumber
\\[3mm]
&& {N}/{L} = \frac{1}{2} \left[ \mathbb{E}(\tau)\lr{ a_{\rm str} +
\frac{{\lambda'}}{b_{\rm str}} } - \mathbb{K}(\tau) \lr{ b_{\rm str} +
\frac{{\lambda'}}{a_{\rm str}} } \right] \, ,
\ea
where $\lambda'=g^2 N_c/(\pi L)^2$ is the so-called BMN coupling~\cite{BMN02}.
The auxiliary variables $\tau$, $a_{\rm str}$ and $b_{\rm str}$ parameterize
classical rotating string solution and verify the relations
\be\label{b-str}
a_{\rm str}=b_{\rm str}/\sqrt{1-\tau}\,,\qquad b_{\rm str} =
\frac{1}{\mathbb{K}(\tau)} \left[ \left(1-\frac{\lambda^\prime}{a_{\rm
str}^2}\right) \left(1 - \frac{\lambda^\prime}{b_{\rm str}^2}\right)
\right]^{-1/2}.
\ee
Their substitution into \re{N/L} yields the anomalous dimension $\gamma_L(N)/L$
as a function of  $N/L$ through parametric dependence of both functions on the
auxiliary modular parameter $\tau$.

Let us apply \re{N/L} to reconstruct the corresponding function $f_L(N)$. In the
$\mathcal{N}=4$ SYM theory, the defining relation \re{f-beta=0} takes the form
\be\label{f-ads-def}
\gamma_L(N) = f_L\left(N+\ft12\gamma_L(N)\right)\,.
\ee
We recall that the shift of the argument in the right-hand side of this relation
takes into account renormalization of the conformal spin of the operators \re{O}
in higher loops. The operators \re{O} have Lorentz spin $N$ and canonical
dimension $N+L$, so that their `bare' conformal spin equals $j=N+\ft12 L$.

We have seen in the previous sections, that for the twist-two operators with
large Lorentz spin $N$ the function $f_{L=2}(N)$ admits asymptotic expansion in
inverse powers of the $SL(2)$ Casimir $J^2=j(j-1)$. In the case under
consideration, an important difference as compared with the twist two is that,
for $\lambda \gg 1$ and ${N}/{\sqrt{\lambda}}$, ${L}/{\sqrt{\lambda}}={\rm
fixed}$, both the Lorentz spin $N$ and the twist $L$ are necessarily large and,
therefore, discussing large $N$ asymptotics of $f_L(N)$ one has to distinguish
two different limits $N\gg L \gg 1$ and $L\gg N \gg 1$. We will see in a moment
that the function $f_L(N)$ has the parity preserving property in the former limit
only. Introducing the scaling variable $\alpha=N/L$ (not to be confused with the
physical coupling \re{alpha}), one finds the quadratic Casimir for the operators
\re{O} as
\be\label{large-J2}
J^2 = L^2 \lr{\alpha+\ft12}^2 \left[1-\frac{1}{L\lr{\alpha+\ft12}}\right] =  L^2
\lr{\alpha+\ft12}^2 \left[1+\mathcal{O}(1/\sqrt{\lambda})\right]\,.
\ee
We expect that in the scaling limit, $\lambda \gg 1$ and ${\alpha}={\rm fixed}$,
the function $f_L(N)$ should admit asymptotic expansion in integer negative
powers of $J^2$, or equivalently in even negative powers of $(\alpha+\ft12)$. To
verify this property, we shall apply Eqs.~\re{N/L} -- \re{f-ads-def} to
reconstruct the large $N$ expansion of $f(N)$ and, then, examine the resulting
series for different values of $\alpha=N/L$.

The coupling constant enters into the right-hand side of both relations in
\re{N/L} through the BMN coupling $\lambda'=g^2 N_c/(\pi L)^2$. Assuming that the
modular parameter $\tau$ admits a regular expansion in powers of $\lambda'$
\footnote{As was shown in Refs.~\cite{BGK06,SS06}, this assumption is only
justified for $\lambda'\ln^2(N/L) < 1$.}
\be\label{b-exp}
\tau = \tau^{(0)} + \lambda^\prime \,\tau^{(1)} + \ldots \, ,
\ee
one finds from the second relation in \re{N/L} for $\lambda^\prime=0$
\be\label{b0}
\alpha+\frac12 =
\frac12\frac{\mathbb{E}(\tau^{(0)})}{\sqrt{1-\tau^{(0)}}\mathbb{K}(\tau^{(0)})}
\,.
\ee
Examining subleading $\sim (\lambda^\prime)^k$ corrections to \re{N/L} one can
express $\tau^{(k)}-$coef\-ficients in terms of the leading one $\tau^{(0)}$.
Then, substitution of \re{b-exp} into the first relation in \re{N/L} yields
asymptotic expansion of the
anomalous dimension in powers of the BMN coupling~\cite{BFST03,BGK06} %
\footnote{The one-loop quantum string corrections to the energy of folded
rotating string \re{E0} break the BMN scaling of the anomalous
dimension~\cite{FTT06}.}
\be
\label{AnomalousDimensionPertExp}
\gamma_L(N) = L\left[ \lambda' \gamma^{(0)}+ {(\lambda')}^2\gamma^{(1)}+ \ldots
\right] \, .
\ee
The first few coefficients are given by
\ba \nonumber
 \gamma^{(0)} \!\!\!&=&\!\!\! \frac{1}2 \mathbb{K}
(\tau^{(0)}) \left[ (2 - \tau^{(0)}) \mathbb{K} (\tau^{(0)}) - 2 \mathbb{E}
(\tau^{(0)}) \right] \, ,
\\
\label{gamma2Diff} \gamma^{(1)} \!\!\!&=&\!\!\! \frac{1}{8} \mathbb{K}^{3}
(\tau^{(0)}) \left[ \lr{ 4 (2 - \tau^{(0)}) \sqrt{1 - \tau^{(0)}} -
(\tau^{(0)})^2} \mathbb{K} (\tau^{(0)}) - 8 \sqrt{1 - \tau^{(0)}} \mathbb{E}
(\tau^{(0)}) \right] \, .
\ea
The relations \re{b0} -- \re{gamma2Diff} define parametric dependence of the
anomalous dimension $\gamma_L(N)$ on the rescaled Lorentz spin $\alpha=N/L$. The
leading $\sim\lambda'$ term in \re{AnomalousDimensionPertExp} coincides with the
one-loop expression for the minimal anomalous dimension of the operators \re{O}
in the thermodynamical limit $L \gg 1$~\cite{BFST03,BGK06}.

Let us combine together the relations \re{f-ads-def} and
\re{AnomalousDimensionPertExp} and determine the function $f_L(N)$. We find that
$f_L(N)$ also admits an expansion in powers of the BMN coupling
\be\label{f-BMN}
f_L(N) = L\left[ \lambda' f^{(0)}+ {(\lambda')}^2f^{(1)}+ \ldots \right]
\ee
with the coefficients related to \re{gamma2Diff} as
\ba \nonumber
f^{(0)} &=& \gamma^{(0)} \, ,
\\ \label{f-ads}
f^{(1)} &=&  \gamma^{(1)} - \frac12 \gamma^{(0)}\frac{\partial_{\tau^{(0)}}
\gamma^{(0)}}{\partial_{\tau^{(0)}}\alpha} =-\frac{1}{8} \mathbb{K}^{4}
(\tau^{(0)}) \lr{\tau^{(0)}}^2\,.
\ea
We observe a remarkable simplification of $f^{(1)}$ as compared to
$\gamma^{(1)}$, Eq.~\re{gamma2Diff}. Let us examine the expressions for
$\gamma_L(N)$ and $f_L(N)$ for different values of $\alpha=N/L$.

\subsection*{$L\gg N\gg 1$}

In this limit, for $\alpha\ll 1$, one finds from \re{b0} that $\alpha = {\frac
{1}{32}}\lr{\tau^{(0)}}^{2}+{\frac {1}{32}}\lr{\tau^{(0)}}^{3}+\ldots$ with
$\tau^{(0)} \ll 1$. Expanding \re{AnomalousDimensionPertExp} and \re{f-BMN} in
powers of $\tau^{(0)}$ one gets the leading term
\be
\gamma^{(0)} = f^{(0)} = \frac12\, {\pi }^{2} \left[ {\alpha}-{\frac
{1}{2}}{{\alpha}} ^{2}+{\frac {3}{8}}{{\alpha}}^{3}+\ldots \right],
\ee
and the first subleading correction
\ba\nonumber
\gamma^{(1)} &=& -\frac18\, {\pi }^{4} \left[{\alpha}+2\,{{\alpha}}^{2}-{ \frac
{11}{4}}{{\alpha}}^{3}+\ldots \right],
\\
f^{(1)} &=& -\frac14\, {\pi }^{4} \left[{\alpha}+{\frac {1}{4}}{{\alpha}
}^{2}-{\frac {3}{8}}{{\alpha}}^{3}+\ldots \right].
\ea
%and to next-to-next-to-leading order
%\ba\nonumber
%\gamma^{(2)} &=& \frac1{16}{\pi }^{6} \left[{\alpha}+4\,{{\alpha}}^{2}+{ \frac
%{5}{8}}{{\alpha}}^{3}+\ldots \right]
%\\
%f^{(2)} &=& {\frac {5}{32}}\,{\pi }^{6}\left[{\alpha}+{\frac {19}{10}
%}{{\alpha}}^{2}-{\frac {6}{5}}{{\alpha}}^{3}+\ldots \right]
%\ea
We observe that the two functions, $\gamma_L(N)$ and $f_L(N)$, vanish for
$\alpha\to 0$. Expansion of $f_L(N)$ involves all powers of $\alpha=N/L$
and does not reveal any symmetry.%
\footnote{The situation is similar to that for the one-loop anomalous dimension
of twist-two operators in the $\mathcal{N}=4$ SYM theory, Eq.~\re{anom-4}. The
anomalous dimension $\gamma_0(N)=4[\psi(N+1)-\psi(1)]$ vanishes for $N=0$, its
expansion around $N=0$ involves all powers of $N$ but its large $N$ expansion
runs in integer powers of $1/(N(N+1))$.}

\subsection*{$N\gg L\gg 1$}

In this limit, for $\alpha\gg 1$, one finds from \re{b0} that $\tau^{(0)} \to 1$.
Denoting $\tau^{(0)}=1-16 z^2$ one obtains from \re{b0} for $z\to 0$
\be\label{a-x}
\alpha+\ft12 = \frac1{z} \left[{\frac {1}{8\ell}}+ {\frac
{2\,{\ell}^{2}-2\,\ell+1}{2\ell^2}}z^2+\mathcal{O} \left( {z}^{4} \right)
\right],
\ee
with $\ell=\ln (1/z)$. One notices that the series inside the square brackets in
the right-hand side of \re{a-x} only involves even powers of $z$. Therefore,
inverting \re{a-x} one obtains $z$ as a series in {\it odd} negative powers of
$(\alpha+\ft12)$ with the expansion coefficients depending on
$\ln(\alpha+\ft12)$.

Replacing $\tau^{(0)}=1-16 z^2$ in \re{gamma2Diff} and \re{f-ads} and expanding
the resulting expressions around $z=0$ one gets the leading term
\be\label{LO-exp}
\gamma^{(0)} = f^{(0)}=  { \ell}^2\left[ {\frac {\ell-2}{2\ell}}+4{\frac
{{\ell}^{2}-\ell+1}{{\ell}^{2}}}z^2+\mathcal{O} \left( {z}^{4} \right) \right],
\ee
as well as the subleading corrections to the anomalous dimension $\gamma_L(N)$,
Eq.~\re{AnomalousDimensionPertExp},
\ba
%\nonumber
%\gamma^{(0)} &=&  { \ell}^2\left[ {\frac {\ell-2}{2\ell}}+4{\frac
%{{\ell}^{2}-\ell+1}{{\ell}^{2}}}z^2+\mathcal{O} \left( {z}^{4} \right) \right]
%\\
\label{gamma_k} \gamma^{(1)} &=&  \ell^4\left[-{\frac {1}{8}}+ {\frac { 2\left(
\ell-2 \right) }{\ell}}{z}+  {\frac { 2\left( \ell+1 \right)}{\ell}} z^2
+\mathcal{O} \left( {z}^{3} \right) \right]\,,
\\ \nonumber
\gamma^{(2)} &=& \ell^6 \left[{\frac {1}{16}}\,{\frac {\ell+1}{\ell-1}}- {\frac {
 \left( 3\,\ell-5 \right) }{2(\ell-1)}} {z}-
{\frac {4 {\ell}^{4}-43{\ell}^{3}+137 {\ell}^{2}-160\ell+61}{ 2\ell\, \left(
\ell-1 \right) ^{2}}}z^2
 +\mathcal{O} \left( {z}^3  \right)
\right]\,,
\ea
and to the function $f_L(N)$, Eq.~\re{f-BMN},
\ba
%\nonumber f^{(0)} &=& {\ell}^2\left[ {\frac {\ell-2}{2\ell}}+4{\frac
%{{\ell}^{2}-\ell+1}{{\ell}^{2}}}z^2+\mathcal{O} \left( {z}^{4} \right) \right]
%\\
\label{f_k} f^{(1)} &=& \ell^4\left[ -{\frac {1}{8}}+ {\frac
{2(\ell+1)}{\ell}}z^2+\mathcal{O} \left( {z}^{4} \right) \right]\,,
\\ \nonumber
f^{(2)} &=& \ell^6\left[{\frac {1}{16}}\,{\frac {\ell+1}{\ell-1}}- {\frac
{{\ell}^{3}+{\ell}^{2}-3}{2\ell\, \left( \ell-1 \right) ^{2}}}z^2+O \left( {z}^
{4} \right)  \right].
\ea
Here as compared with \re{gamma2Diff} and \re{f-ads} we also included the
next-to-subleading correction to the functions \re{AnomalousDimensionPertExp} and
\re{f-BMN}.

As follows from \re{LO-exp}, expansion of the leading order term,
$\gamma^{(0)}=f^{(0)}$, runs in even powers of $z\sim (\alpha+\ft12)^{-1}$. Going
over to subleading terms, Eqs.~\re{gamma_k} and \re{f_k}, one finds that
$\gamma^{(k)}$ and $f^{(k)}$ (with $k=1,2,...$) coincide at the level of $\sim
z^0$ corrections but deviate otherwise. Remarkably enough, the functions
$f^{(k)}$ do not contain odd powers of $z$ while they are present in
$\gamma^{(k)}$. Together with \re{a-x} this implies that, in agreement with our
expectations, the large $\alpha$ expansion of the function $f_L(N)$ runs in even
negative powers of $(\alpha+\ft12)$ only.

The relations \re{gamma_k} and \re{f_k} were derived under tacit assumption that
the coefficient functions entering \re{AnomalousDimensionPertExp} and \re{f-BMN}
are uniformly bounded functions of $\alpha=N/L$. Examining the expressions
\re{LO-exp}, \re{gamma_k} and \re{f_k} one finds that for $N\gg L$, or
equivalently $\ell \to \infty$, this assumption is justified provided that
$\lambda' \ell^2 \ll 1$, or equivalently $({\lambda}/{L^2})\ln^2({N}/{L}) \ll 1$.
In the opposite limit, for $\lambda' \ell^2 \gg 1$, one finds that the leading
corrections $\sim \lr{\lambda' \ell^{2}}^n$ to \re{f-BMN} can be resummed to all
orders~\cite{BGK06}
\ba\label{log-scal}
f_L(N) &=&  L \left[\sqrt{1+ \lambda' \ell^2}-1\right] + \ldots \sim
\frac{\sqrt{g^2 N_c}}{\pi}\,\ln N + \ldots\,,
\ea
where $\lambda'=g^2 N_c/(\pi L)^2$ and ellipses denote terms suppressed by powers
of $\ell \sim \ln (\alpha+\ft12)$. We notice that despite the fact that each term
of the expansion is suppressed by inverse power of $L$, the resummed expression
does not depend on the twist $L$ and exhibits a well-known logarithmic
scaling~\cite{GKP02}. For the twist-two anomalous dimensions, this scaling
behaviour arises for $N\gg 1$. Specific feature of operators of higher twist
$L\gg 1$ is that the logarithmic scaling \re{log-scal} sets up for much larger
values of the Lorentz spin such that
\be
\frac{\lambda}{L^2}\ln^2\frac{N}{L} \gg 1\,.
\ee
In this kinematical region, analysis % of analytical properties of the function $f(N)$
becomes more involved due to a complicated form of the defining relations
\re{N/L} and \re{b-str}.

There is however a much simpler way to understand analytical properties of the
function $f_L(N)$. Let us consider the relation \re{f-ads-def} and rewrite it
using \re{N/L} as
\be\label{f-eq}
f_L \lr{\frac{x(\tau)L}2} = L\left[  \mathbb{K}(\tau)\, b_{\rm str}(\tau) -
\mathbb{E}(\tau) \frac{{\lambda'}}{b_{\rm str}(\tau)} - 1\right]\,,
\ee
where $x$ is a function of the modular parameter $\tau$
\be
x(\tau) = \mathbb{E}(\tau)\, a_{\rm str}(\tau) - \mathbb{K}(\tau)
\frac{{\lambda'}}{a_{\rm str}(\tau)} - 1\,,
\ee
with $a_{\rm str}=b_{\rm str}/\sqrt{1-\tau}$ and $b_{\rm str}$ defined in
\re{b-str}. We expect that for large Lorentz spin $N$ the asymptotic expansion of
the function $f_L(N)$ should run in integer negative powers of $J^2 = j^2
\big[1+\mathcal{O}(1/\sqrt{\lambda})\big]$, Eq.~\re{large-J2}, with $j=N+\ft12 L$
being the conformal spin. Substituting $N$ with $\ft12{x(\tau)L}$ as in
\re{f-eq}, one finds that the conformal spin takes the form
\be\label{j-ads}
j = \frac{L}2\left(x(\tau)+1\right)=\frac{L}2\left[ \mathbb{E}(\tau)\, a_{\rm
str}(\tau) - {\lambda'}\frac{\mathbb{K}(\tau)}{a_{\rm str}(\tau)}\right]\,.
\ee
To verify the parity preserving property \re{prop} for $f_L$ it is sufficient to
check that the right-hand side of \re{f-eq} stays invariant under transformation
$j\to -j$ modulo `$i\pi$'--contribution from $\ln j$ terms. Let us show that this
transformation corresponds to going around $\tau=1$ in the complex $\tau-$plane
in the right-hand side of \re{j-ads}. According to \re{a-x}, the limit $N\gg L$
corresponds to $\tau\to 1$. Let us introduce a new variable $\tau=1-16 z^2$ and
examine the relations \re{f-eq} and \re{j-ads} for $z\to 0$. One first verifies
that the elliptic functions $\mathbb{K}(\tau)$ and $\mathbb{E}(\tau)$ have
logarithmic branch cuts that start at $\tau=1$. Their expansion around $z=0$ runs
in even powers of $z^2$ with the coefficients linear in $\ln z$. Examining the
relations \re{b-str} one finds that $b_{\rm str}$ has similar analytical
properties around $\tau=1$, while $a_{\rm str}$ has additional square-root
singularity $a_{\rm str}=b_{\rm str}/\sqrt{1-\tau}=b_{\rm str}/(4z)$. Therefore,
going around $\tau=1$ in the complex $\tau-$plane, or equivalently replacing
$z\to -z$, one finds that, modulo contribution from logarithmic cuts, the
functions $\mathbb{K}(\tau)$, $\mathbb{E}(\tau)$ and $b_{\rm str}$ stay invariant
while $a_{\rm str}$ changes a sign. Going back to \re{f-eq} and \re{j-ads}, we
conclude that the function $f_L(N)$ is indeed invariant under $j\to -j$ and,
therefore, it has the parity preserving property.

\section{Conclusions}

Anomalous dimensions of Wilson operators with large Lorentz spin $N$ admit an
asymptotic expansion in inverse powers of $N$. The leading term of the expansion
scales logarithmically with the spin and is related to the universal cusp
anomalous dimension. The subleading, power suppressed, corrections to the
anomalous dimensions are not universal and depend on the operator under
consideration. Motivated by findings of Refs.~\cite{DKT95,MVV04,VMV04,DMS}, we
argued in this paper that the subleading corrections satisfy infinite number of
coupled relations. They allow one to reconstruct the corrections to the anomalous
dimensions suppressed by odd powers of $J=\sqrt{N(N+1)}$ in terms of the
coefficients accompanying smaller even powers of $J$ to less number of loops.
Remarkably enough, these relations take the same form for different twist-two
operators even though they relate to each other quantities that are operator
dependent.

We demonstrated that the above-mentioned properties of the subleading corrections
naturally follow from the parity preserving property of the scaling function
$f(N)$. This function is related  to the anomalous dimension $\gamma(N)$ through
the functional relation \re{f-def} which generalizes similar relation proposed in
\cite{DKT95,DMS}. It incorporates the constraints imposed by the conformal
invariance and takes into account the conformal symmetry breaking corrections to
the anomalous dimensions due to nonzero beta-function. We would like to stress
that the functional relation \re{f-def} taken alone does not tell us much about
the properties of the anomalous dimensions unless it is supplemented by
additional condition for the function $f(N)$. The latter is just the parity
preserving property of the scaling function $f(N)$, Eq.~\re{prop}.

%Being combined together, the relations \re{f-def} and \re{prop} allow us to
%express the odd-parity corrections $\sim 1/J^{2n+1}$ (with $n=1,2,\ldots$) to the
%anomalous dimensions $\gamma(N)$ in terms of the function $f(N)$.
It is important to keep in mind that the anomalous dimensions $\gamma(N)$ and the
scaling functions $f(N)$ have (Regge) singularities at negative $N$, and
therefore they cannot be invariant under the transformation $N\to -1-N$, or
equivalently $J\to -J$. The parity preserving property \re{prop} only holds true
for each individual term in the asymptotic series for $f(N)$ and the Regge
singularities manifest themselves through a factorial growth of the expansion
coefficients at higher orders in $1/J$. It is interesting to note that the parity
respecting relation is not a unique feature of gauge theories. Using the
four-loop result for the twist-two anomalous dimension in scalar
$\phi^4-$theory~\cite{DGM97}, we verified that the corresponding scaling function
$f(N)$ has the same form as \re{prop} with the only difference that the leading
$\ln \bar J$ term is missing.

In a gauge theory with nonzero beta-function, both the anomalous dimensions and
the scaling functions are scheme-dependent. In our analysis, we employed the
dimensional regularization/reduction scheme because the beta-function
contribution to the anomalous dimension can be described in this scheme by going
over to the `critical' number of space-time dimensions $d_{\rm
cr}=4-2\varepsilon_{\rm cr}$ in which the underlying gauge theory becomes
conformal~\cite{V}. The change of the scheme corresponds to a finite
renormalization of the Wilson operator, say $\mathcal{O}_N^{\scriptscriptstyle
(\rm R)}(0) = Z(N) \mathcal{O}_N(0)$. Then, the anomalous dimension in a new
scheme reads $\gamma_S^{\scriptscriptstyle (\rm R)}(N) = \gamma_S(N)
-\beta(\lambda) \partial_{\ln\lambda}\ln Z(N)$, so that the properties of
$\gamma_S^{\scriptscriptstyle (\rm R)}(N)$ at large $N$ are now governed by the
scheme-dependent factor $Z(N)$. Substituting $\gamma_S^{\scriptscriptstyle (\rm
R)}(N)$ into \re{f-hat} one can reconstruct the corresponding scaling function
$f^{\scriptscriptstyle (\rm R)}(N) = f(N) + \mathcal{O}(\beta(\lambda))$ and
verify that it does not satisfy \re{prop} for generic $Z(N)$. However, imposing
\re{prop} as a condition for $Z(N)$, one can define the renormalization scheme in
which the subleading corrections to the anomalous dimensions still verify the
relations discussed above.

In this paper, we used explicit expressions for the two-loop and three-loop
anomalous dimensions of twist two in QCD and in SYM theories to verify the parity
preserving relation \re{prop}. One may wonder whether the same property holds
true for the anomalous dimensions of higher twist. %We demonstrated that for the
%quasipartonic operators of twist $L$ the scaling function is related to the
%anomalous dimension through the relation \re{f-def}.
Going over to higher twists, one finds that, in general, there are few operators
carrying the same conformal spin. In distinction with the twist two, the size of
their mixing matrix grows with $N$ and its eigenvalues can not be written in
explicit form as functions of $N$. This makes the large $N$ analysis of
higher-twist anomalous dimensions extremely difficult. Tremendous simplification
occurs for the special subclass of quasipartonic operators due to hidden
integrability symmetry of their mixing
matrix in the planar ('t Hooft) limit. %
\footnote{Integrability phenomenon in four-dimensional Yang-Mills theories has
been first discovered in Refs.~\cite{L93,FK95} in context of high-energy (Regge)
asymptotics of the scattering amplitudes.} In QCD and in SYM theories with
$\mathcal{N}=0,1,2$ supercharges, integrability arises for the so-called
maximally helicity operators~\cite{BDM98,BDKM04} while in the $\mathcal{N}=4$
theory it gets extended to a larger class of operators~\cite{L98}. Making use of
the integrability, one can work out a systematic large $N$ expansion of
higher-twist anomalous dimensions in the planar approximation. In this way, one
finds~\cite{K96,BGK06} that for the so-called single-trace operators built from
$L$ fundamental fields belonging to the adjoint representation of the $SU(N_c)$
group and carrying the conformal spin $s$ ($s=\ft12$ for scalars, $s=1$ for
gaugino and $s=\ft32$ for gauge strength tensor), the asymptotic expansion of the
one-loop anomalous dimension runs in integer negative powers of the conformal
Casimir $J^2=(N+Ls)(N+Ls-1)$. It would be interesting to extend analysis to
higher orders in the coupling constant and to verify the parity respecting
relation for the scaling function $f_L(N)$ in higher loops. The gauge/string
correspondence suggests that it should hold true at least for the minimal
anomalous dimension in the strong coupling regime in the $\mathcal{N}=4$ SYM
theory for higher twists and large spins such that $N\gg L \gg 1$.

\paragraph{Note added:}
When this paper was in writing we learned from Yuri Dokshitzer and Pino
Marchesini that they studied the reciprocity respecting equation \re{DMS} in the
$\mathcal{N}=4$ SYM theory and analyzed the properties of the corresponding
three-loop scaling function $f(N)$ in important kinematical limits. Our results
agree with theirs~\cite{DM06} as far as the parity respecting property \re{prop}
and the structure of the large $N$ expansion of $f(N)$, Eq.~\re{OK}, are
concerned.

\section*{Acknowledgements}

We are most grateful to S.~Moch for collaboration at an early stage of the
project. We would like to thank Yu.~Dokshitzer, E.~Gardi, A.~Gorsky, A.~Manashov,
G.~Marchesini and A.~Mueller for interesting discussions and J.~Gracey for useful
correspondence. The work was supported in part by the French Agence Nationale de
la Recherche under grant ANR-06-BLAN-0142-02.

\appendix

\setcounter{section}{0} \setcounter{equation}{0}
\renewcommand{\theequation}{\Alph{section}.\arabic{equation}}

\section{Large $N$ expansion of the anomalous dimensions}

In this appendix we present expressions for the large $N$ expansion of various
twist-two anomalous dimensions $\gamma_S(N)$ and the corresponding functions
$f(N)$. They were obtained in collaboration with S.~Moch. The anomalous
dimensions are given by a series \re{gamma-exp} in the coupling constant
$\lambda$ defined in \re{couplage}. At large $N$ their leading asymptotic
behaviour is
\be
\gamma_S(N) = 2 \Gamma_{\rm cusp}(\lambda) \ln \bar N + \mathcal{O}(N^0)\,,
\ee
with the cusp anomalous dimension $\Gamma_{\rm cusp}(\lambda)$ depending on the
gauge theory under consideration. Perturbative expansion of $\gamma_S(N)$ can be
simplified by using the `physical' coupling~\cite{CMW91}, $\alpha = \ft12
\Gamma_{\rm cusp}(\lambda)$. Then, the relations \re{gamma-exp} and \re{f-exp}
can be rewritten in terms of this coupling as
\ba\nonumber
\gamma_S(N) &=& \alpha \lr{4\ln \bar N + \gamma_{0,\rm ph}}
+\alpha^2\gamma_{1,\rm ph} +\alpha^3\gamma_{2,\rm ph} + \mathcal{O}(\alpha^4)\,,
\\[2mm] \label{f-ph}
f(N)&=&\alpha \lr{2\ln \bar J^2 + f_{0,\rm ph}} +\alpha^2f_{1,\rm ph}
+\alpha^3f_{2,\rm ph} + \mathcal{O}(\alpha^4)\,,
\ea
where $\bar N=N\e^{\gamma_{\rm \scriptscriptstyle E}}$, $\bar J=J\e^{\gamma_{\rm
\scriptscriptstyle E}}$ and $J^2=N(N+1)$ is the conformal Casimir. Here, the
$\gamma_{\rm ph}-$coefficients are given by series in integer negative powers of
$N$ while the $f_{\rm ph}-$coefficients admit expansion in {\it even} negative
powers of $J$ only, Eq.~\re{prop}.

The multi-loop twist-two anomalous dimensions $\gamma_S(N)$ are expressed in
terms of generalized harmonic (Euler-Zagier) sums. In QCD, they also depend on
the $SU(N_c)$ color and $SU(n_f)$ flavour factors and their explicit expressions
are cumbersome~\cite{MVV04,VMV04}. In the $\mathcal{N}=4$ SYM, to three-loop
accuracy, the perturbative coefficients in \re{f-ph} are $N_c-$independent and,
as a consequence, they are much simpler as compared to the QCD case. For the sake
of simplicity, we shall present below the explicit expressions \re{f-ph} in the
$\mathcal{N}=4$ SYM theory and outline a difference as compared with the QCD
expressions. In $\mathcal{N}=4$ SYM the anomalous dimensions of all twist-two
operators are given by the same, universal function of $N$~\cite{KLOV04}. To
three-loop accuracy, the cusp anomalous dimension, or equivalently the physical
coupling constant is given by \re{alpha}. Expanding the three-loop result of
Ref.~\cite{KLOV04} at large $N$ and matching it with the first relation in
\re{f-ph} one finds
\ba\nonumber
\gamma_{0,{\rm ph}} &=&
2\,\b{{N}^{-1}}-\ft13\,\b{{N}^{-2}}+\ft1{30}\,\b{{N}^{-4}}+\mathcal{O}
 \left(\b{ {N}^{-6}} \right)\,,
\\[2mm]\nonumber
\gamma_{1,{\rm ph}} &=& -6\,\zeta_3 +\lr{8\,\ln\bar N}\b{{N}^{-1}}+{ \lr
{6-4\,\ln\bar N }\b{{N}^{-2}}}
\\[1mm]\nonumber
&+&{\lr{\ft43\,\ln\bar N -\ft{14}3}\b{{N}^{-3}}}+4\,\b{{N}^{-4}}+\mathcal{O}
\left( \b{{N}^{-5}}
 \right)\,,
\\[2mm]\nonumber
\gamma_{2,{\rm ph}} &{=}& \lr{-\ft43\,{\pi }^{2}\zeta_3 +20\,\zeta_5} - 12\,{
{\zeta_3 }\,\b{{N}^{-1}}}
\\[1mm]\nonumber
&{+}&{\lr {-8\, ( \ln\bar N) ^{2}+\lr{-\ft23\,{\pi }^{2}+16}\ln\bar N
  +\ft23\,{\pi }^{2}+6\,\zeta_3
}\b{ {N}^{-2}}}
\\\nonumber
&+& {\lr{8\, \left( \ln\bar N \right) ^{2}+\lr{\ft23\,{\pi }^{2}-32}\ln\bar N +
12  -{\pi }^{2}-2\,\zeta_3}\b{{N}^{-3}}}
\\ \label{g-ph}
&+& \left( -4\, (\ln\bar N)^{2}+{\ft{110}{3}}\,\ln\bar N-{\ft{70}{3}} +{\ft
{13}{18}}\,{\pi }^{2} \right) \b{{N}^{-4 }}+\mathcal{O} \left( \b{{N}^{-5}}
\right)\,.
\ea
One evaluates the corresponding function $f(N)$ with a help of Eqs.~\re{f-hat}
and \re{f-tilde} and, then, expands it in inverse powers of $J^2=N(N+1)$ to get
from \re{f-ph}
\ba \nonumber
f_{0,{\rm ph}}&=& \ft23\,\b{{J}^{-2}}-\ft2{15}\,\b{{J}^{-4}}
+\mathcal{O} \left( \b{{J}^{-6}} \right)\,,
%+{\ft{16}{315 }}\,\b{{J}^{-6}}-{\ft {4}{105}}\,\b{{J}^{-8}}+\mathcal{O} \left(
%\b{{J}^{-10}} \right)
\\[2mm] \nonumber
f_{1,{\rm ph}}&=& -6\,\zeta_3 +2\,\b{{J}^{-2}}+\b{{J}^{-4}}
+\mathcal{O} \left( \b{{J}^{-6}} \right)\,,
%-{\ft {22}{3}}\,\b{{J}
%^{-6}}+{\ft {130}{3}}\,\b{{J}^{-8}}+\mathcal{O} \left(\b{{J}^{-10}} \right)
\\[2mm] \nonumber
f_{2,{\rm ph}}&=&\lr{20\,\zeta_5 -\ft43\,{\pi }^{2}\zeta_3} +{ \lr {-\ft23\,{\pi
}^{2}\ln\bar J +\ft23\,{\pi }^{2}}\b{{J}^{-2} }}
\\[1mm] \label{ff-ph}
&+&{\lr{\lr{6 +\ft23\,{\pi }^{2}}\ln\bar J -\ft49\,{\pi }^{2}-2}\b{{J}^{-4}}}
+\mathcal{O} \left( \b{{J}^{-6}} \right)\,.
%\\[2mm]
% &+& \left( -\lr{32+\ft43\,{\pi }^{2}}{\ln\bar J} +\ft{11}3+{\ft {16}{45}}
%\,{\pi }^{2} \right) \b{{J}^{-6}}+ \left(\lr{222+\ft{16}3\,{\pi }^{2}}\ln\bar
%J-{\ft{449}{945}}\,{\pi }^{2} -{ \ft {406}{5}} \right) \b{{J}^{-8}}+\mathcal{O}
%\left(\b{{J}^{-10}} \right)
\ea
Examining these relations, one observes that for the anomalous dimension
$\gamma_S(N)$
%, in the large $N$ expansion of $\gamma(N)$,
the coefficients in front of negative powers of $N$ in the right-hand side of
\re{g-ph} are enhanced by powers of $\ln\bar N$ with the leading term
$\gamma_{n,{\rm ph}}\sim \lr{\ln\bar N/N}^n$ (for $n=1,2$). For the function
$f(N)$ one finds that, in agreement with the parity preserving relation
\re{prop}, its expansion runs in negative powers of $J^2$. Remarkably enough, the
one-loop and two-loop coefficients, $f_{0,{\rm ph}}$ and $f_{1,{\rm ph}}$
respectively, do not involve logarithmically enhanced terms but they reappear
starting from three loops $f_{2,{\rm ph}}\sim \ln\bar J/J^2 \sim \ln\bar N/N^2$
(to be compared with $\gamma_{2,{\rm ph}}\sim (\ln\bar N)^2/N^2$).

Repeating similar analysis for the twist-two operators in QCD, we found that, in
spite of the fact that the large $N$ expressions for the functions $\gamma(N)$
and $f(N)$ are much more cumbersome, the $\sim \ln \bar N$ and $\sim\ln\bar J$
terms inside the $\gamma_{{\rm ph}}-$ and $f_{{\rm ph}}-$coefficients have the
same properties as in the $\mathcal{N}=4$ theory for the following anomalous
dimensions: nonsinglet unpolarized quark operators to three loops~\cite{MVV04},
quark transversity operator to two loops~\cite{V97} and linearly polarized gluon
operator to two loops~\cite{V98}.

For the singlet (un)polarized twist-two anomalous dimensions in QCD, the analysis
becomes more cumbersome since one has to solve the characteristic equation
\re{ch-eq} and, then, use its solutions, $\gamma_q(N)$ and $\gamma_g(N)$, to
define the corresponding functions $f_q(N)$ and $f_g(N)$ with a help of
\re{f-hat} and \re{f-tilde}. As in \re{X,Y}, it %Sect.~2.2, it
is more advantageous to analyze the properties of the functions $\widehat
f_q(N)+\widehat f_g(N)$ and $\widehat f_q(N)\widehat f_g(N)$.
%
%the functions\\[-1mm]
%\be
%\widehat X_f(N) = \widehat f_q(N)+\widehat f_g(N)\,,\qquad \widehat Y_f(N) =
%\widehat f_q(N)\widehat f_g(N)\,,
%\ee\\[-1mm]
%which are analogous to \re{X,Y}.
Making use of \re{f-hat} it is straightforward
to verify that
\ba \nonumber
\widehat f_q +\widehat f_g  &=& \widehat X - \ft14 \lr{\widehat X^2-2\widehat
Y}'+\ft1{24}\lr{\widehat X^3 -3 \widehat X\widehat Y}''
+\mathcal{O}(\lambda^4)\,,
\\ \label{rel4}
\widehat f_q \widehat f_g &=& \widehat Y \left[1-\ft12 \widehat X'+\ft14
\big({\widehat X'}\big)^2 +\ft18 \widehat X \widehat X'' - \ft18 \widehat
Y''\right]+\mathcal{O}(\lambda^5)\,,
\ea
where the notation was introduced for the functions
\ba \nonumber
\widehat X(N) &=& \widehat\gamma_q + \widehat\gamma_g = X -2\beta(\lambda)
\,,\qquad
\\[2mm]
\widehat Y(N) &=& \widehat\gamma_q\, \widehat\gamma_g = Y - X
\beta(\lambda) + [\beta(\lambda)]^2\,,
\ea
with $X$ and $Y$ given in \re{X,Y}. In the right-hand side of these relations,
the $N-$dependence of all functions (except of $\beta(\lambda)$) is tacitly
assumed and prime denotes a derivative with respect to $N$. We verified by
explicit calculation that for the three-loop singlet unpolarized anomalous
dimensions~\cite{VMV04} and the two-loop singlet polarized anomalous
dimensions~\cite{MN95}, the right-hand side of both relations in \re{rel4} has
asymptotic expansion in integer powers of $1/J^2$. This implies that the
corresponding functions $f_q$ and $f_g$ (or equivalently $\widehat f_q$ and
$\widehat f_g$) possess the parity preserving property \re{prop}. The resulting
expressions can be described as follows. The off-diagonal elements of the mixing
matrix scale as $\gamma^{gq},\,\gamma^{qg} \sim 1/J$ (see Eq.~\re{anom-sing}) and
they affect $f_q$ and $f_g$ at the level of $\mathcal{O}(J^{-2})$ corrections
only. As a result, for both functions, $f_q$ and $f_g$, the corresponding $f_{\rm
ph}-$coefficients \re{f-ph} take the form \re{ff-ph} with the only difference
that the coefficients in front of $J^{-2}, J^{-4}, \ldots$ receive an additional
contribution given by (an infinite) series in $1/\ln \bar J$.

%%%%%%%%%%%%%%%%%%%%%%%%%%%%%%%%%%%%%%%%%%%%%%%%%%%%%%%%%%%%%%%%%%%%%


\begin{thebibliography}{99}
%%%%%%%%%%%%%%%%%%%%%%%%%%%%%%%%%%%%%%%%%%%%%%%%%%%%%%%%%%%%%%%%%%%%%

\bibitem{Mal97}
J.M. Maldacena, Adv. Theor. Math. Phys. 2 (1998) 231;\\
%%CITATION = HEP-TH 9711200;%%
%
S.S. Gubser, I.R. Klebanov, A.M. Polyakov,
Phys. Lett. B 428 (1998) 105;\\
%%CITATION = HEP-TH 9802109;%%
%
E. Witten, Adv. Theor. Math. Phys. 2 (1998) 253.
%%CITATION = HEP-TH 9802150;%%

\bibitem{MVV04}
S. Moch, J.A.M. Vermaseren, A. Vogt, Nucl. Phys. B 688 (2004) 101.
%%CITATION = HEP-PH 0403192;%%

\bibitem{VMV04}
A. Vogt, S. Moch, J.A.M. Vermaseren,
%``The three-loop splitting functions in QCD: The singlet case,''
Nucl.\ Phys.\ B {691} (2004) 129. %[arXiv:hep-ph/0404111].
%%CITATION = HEP-PH 0404111;%%

\bibitem{Kor88}
G.P. Korchemsky,
Mod. Phys. Lett. A 4 (1989) 1257; \\
%%CITATION = MPLAE,A4,1257;%%
%
G.P. Korchemsky, G. Marchesini, Nucl. Phys. B 406 (1993) 225.
%%CITATION = HEP-PH 9210281;%%

\bibitem{DMS}
Yu.L. Dokshitzer, G. Marchesini, G.P. Salam,
%``Revisiting parton evolution and the large-x limit,''
Phys.\ Lett.\ B {634} (2006) 504; %[arXiv:hep-ph/0511302].
%%CITATION = HEP-PH 0511302;%%
\\
G. Marchesini,
%``Relating small Feynman and Bjorken x,''
arXiv:hep-ph/0605262.
%%CITATION = HEP-PH 0605262;%%

\bibitem{DKT95}
Yu.L. Dokshitzer, V.A. Khoze, S.I. Troian,
%``Specific features of heavy quark production. LPHD approach to heavy
%particle spectra,''
Phys.\ Rev.\ D {53} (1996) 89. % [arXiv:hep-ph/9506425].
%%CITATION = HEP-PH 9506425;%%

\bibitem{BKM06}
A.V. Belitsky, G.P. Korchemsky, D. M\"uller,
%``Towards Baxter equation in supersymmetric Yang-Mills theories,''
arXiv:hep-th/0605291.
%%CITATION = HEP-TH 0605291;%%

\bibitem{GL72a}
V.N. Gribov, L.N. Lipatov,
%``Deep Inelastic E P Scattering In Perturbation Theory,''
Sov.\ J.\ Nucl.\ Phys.\  {15} (1972) 438.
%  [Yad.\ Fiz.\  {\bf 15} (1972) 781].
%%CITATION = SJNCA,15,438;%%

\bibitem{AP77}
G. Altarelli, G. Parisi,
%``Asymptotic Freedom In Parton Language,''
Nucl.\ Phys.\ B {126} (1977) 298.
%%CITATION = NUPHA,B126,298;%%

\bibitem{D77}
Yu.L. Dokshitzer,
%``Calculation Of The Structure Functions For Deep Inelastic Scattering And E+
%E- Annihilation By Perturbation Theory In Quantum Chromodynamics. (In
%Russian),''
Sov.\ Phys.\ JETP {46} (1977) 641.
%%CITATION = SPHJA,46,641;%%

\bibitem{JS82}
R.L. Jaffe, M. Soldate,
%``Twist Four In Electroproduction: Canonical Operators And Coefficient
%Functions,''
Phys.\ Rev.\ D {26} (1982) 49.
%%CITATION = PHRVA,D26,49;%%

\bibitem{CTEQ}
R. Brock {\it et al.}  [CTEQ Collaboration],
%``Handbook of perturbative QCD: Version 1.0,''
Rev.\ Mod.\ Phys.\  {67} (1995) 157.
%%CITATION = RMPHA,67,157;%%

\bibitem{GL72}
V.N. Gribov, L.N. Lipatov,
%``E+ E- Pair Annihilation And Deep Inelastic E P Scattering In Perturbation
%Theory,''
Sov.\ J.\ Nucl.\ Phys.\  {15} (1972) 675.
%%CITATION = SJNCA,15,675;%%

\bibitem{DLY69}
S.D. Drell, D.J. Levy, T.M. Yan,
%``A Theory Of Deep Inelastic Lepton - Nucleon Scattering And Lepton Pair
%Annihilation Processes. I,''
Phys.\ Rev.\  {187} (1969) 2159;
%%CITATION = PHRVA,187,2159;%%
%``A Theory Of Deep Inelastic Lepton - Nucleon Scattering And Lepton Pair
%Annihilation Processes. 3. Deep Inelastic Electron - Positron Annihilation,''
Phys.\ Rev.\ D {1} (1970) 1617.
%%CITATION = PHRVA,D1,1617;%%

\bibitem{CFP80}
G. Curci, W. Furmanski, R. Petronzio,
%``Evolution Of Parton Densities Beyond Leading Order: The Nonsinglet Case,''
Nucl.\ Phys.\ B {175} (1980) 27.
%%CITATION = NUPHA,B175,27;%%

\bibitem{FP80}
W. Furmanski, R. Petronzio,
%``Singlet Parton Densities Beyond Leading Order,''
Phys.\ Lett.\ B {97} (1980) 437.
%%CITATION = PHLTA,B97,437;%%

\bibitem{FKL81}
E.G. Floratos, C. Kounnas, R. Lacaze,
%``Higher Order QCD Effects In Inclusive Annihilation And Deep Inelastic
%Scattering,''
Nucl.\ Phys.\ B {192} (1981) 41;
%%CITATION = NUPHA,B192,417;%%
Phys.\ Lett.\ B {98} (1981) 89,
%%CITATION = PHLTA,B98,89;%%
285.
%%CITATION = PHLTA,B98,285;%%

\bibitem{SV96}
M. Stratmann, W. Vogelsang,
%``Next-to-leading order evolution of polarized and unpolarized  fragmentation
%functions,''
Nucl.\ Phys.\ B {496} (1997) 41. % [arXiv:hep-ph/9612250].
%%CITATION = HEP-PH 9612250;%%

\bibitem{SV01}
M. Stratmann, W. Vogelsang,
%``Next-to-leading order QCD evolution of transversity fragmentation
%functions,''
Phys.\ Rev.\ D {65} (2002) 057502. % [arXiv:hep-ph/0108241].
%%CITATION = HEP-PH 0108241;%%

\bibitem{BRN00}
J. Blumlein, V. Ravindran, W.L. van Neerven, Nucl.\ Phys.\ B 586 (2000) 349.

\bibitem{Mue83}
A.H. Mueller,
%``Multiplicity And Hadron Distributions In QCD Jets. 2. A General Procedure
%For All Nonleading Terms,''
Nucl.\ Phys.\ B {228} (1983) 351.
%%CITATION = NUPHA,B228,351;%%


\bibitem{MMV06}
A. Mitov, S. Moch, A. Vogt,
%``Next-to-next-to-leading order evolution of non-singlet fragmentation
%functions,''
Phys.\ Lett.\ B {638} (2006) 61.
%  [arXiv:hep-ph/0604053].
%%CITATION = HEP-PH 0604053;%%

\bibitem{BFLK85}
A.P. Bukhvostov, G.V. Frolov, L.N. Lipatov, E.A. Kuraev,
%``Evolution Equations For Quasi - Partonic Operators,''
Nucl.\ Phys.\ B {258} (1985) 601.
%%CITATION = NUPHA,B258,601;%%

\bibitem{BKM03}
V.M. Braun, G.P. Korchemsky, D. M\"uller,
%``The uses of conformal symmetry in QCD,''
Prog.\ Part.\ Nucl.\ Phys.\  {51} (2003) 311. % [arXiv:hep-ph/0306057].
%%CITATION = HEP-PH 0306057;%%

\bibitem{O81}
T. Ohrndorf,
%``Constraints From Conformal Covariance On The Mixing Of Operators Of Lowest
%Twist,''
Nucl.\ Phys.\ B {198} (1982) 26.
%%CITATION = NUPHA,B198,26;%%

\bibitem{M93}
D. M\"uller,
%``Conformal constraints and the evolution of the nonsinglet meson
%distribution amplitude,''
Phys.\ Rev.\ D {49} (1994) 2525;
%%CITATION = PHRVA,D49,2525;%%
%D.~Mueller,
%``Restricted conformal invariance in QCD and its predictive power for
%virtual two-photon processes,''
Phys.\ Rev.\ D {58} (1998) 054005; % [arXiv:hep-ph/9704406].
%%CITATION = HEP-PH 9704406;%%
\\
A.V. Belitsky, D. M\"uller,
%``Broken conformal invariance and spectrum of anomalous dimensions in
%{QCD},''
Nucl.\ Phys.\ B {\bf 537} (1999) 397. % [arXiv:hep-ph/9804379].
%%CITATION = HEP-PH 9804379;%%

\bibitem{BKM04}
A.V. Belitsky, G.P. Korchemsky, D. M\"uller, Phys. Rev. Lett. 94 (2005) 151603;
%%CITATION = HEP-TH 0412054;%%
%
Nucl. Phys. B 735 (2006) 17.
%%CITATION = HEP-TH 0509121;%%
%

\bibitem{VPK}
A.N. Vasil'ev, Y.M. Pis'mak, Yu.R. Khonkonen,
%``Simple Method Of Calculating The Critical Indices In The 1/N Expansion,''
Theor.\ Math.\ Phys.\  {46} (1981) 104;
%%CITATION = TMPHA,46,104;%%
% A.N.~Vasiliev, Yu.M.~Pismak, Yu.R.~Khonkonen,
%``1/N Expansion: Calculation Of The Exponents Eta And Nu In The Order 1/N**2
%For Arbitrary Number Of Dimensions,''
%Theor.\ Math.\ Phys.\
{47} (1981) 465;
%%CITATION = TMPHA,47,465;%%
%A.~N.~Vasiliev, Yu.~M.~Pismak and Yu.~R.~Khonkonen,
%``1/N Expansion: Calculation Of The Exponent Eta In The Order 1/N**3 By The
%Conformal Bootstrap Method,''
%Theor.\ Math.\ Phys.\
{50} (1982) 127.
%%CITATION = TMPHA,50,127;%%

\bibitem{V}
A.N. Vasil'ev, {\it The Field Theoretic Renormalization Group in Critical
Behavior Theory and Stochastic Dynamics}, Chapman \& Hall/CRC, 2004.

\bibitem{GKP02}
S.S. Gubser, I.R. Klebanov, A.M. Polyakov,
%``A semi-classical limit of the gauge/string correspondence,''
Nucl.\ Phys.\ B {636} (2002) 99. % [arXiv:hep-th/0204051].
%%CITATION = HEP-TH 0204051;%%

\bibitem{FT03}
S. Frolov, A.A. Tseytlin, J. High Ener. Phys. 0206 (2002) 007.
%%CITATION = HEP-TH 0204226;%%

\bibitem{V98}
W. Vogelsang,
%``Q**2 evolution of spin-dependent parton densities,''
Acta Phys.\ Polon.\ B {29} (1998) 1189. % [arXiv:hep-ph/9805295].
%%CITATION = HEP-PH 9805295;%%

\bibitem{MN95}
R. Mertig, W.L.~van Neerven,
%``The Calculation Of The Two Loop Spin Splitting Functions P(Ij)(1)(X),''
Z.\ Phys.\ C {70} (1996) 637; % [arXiv:hep-ph/9506451].
%%CITATION = HEP-PH 9506451;%%
\\
%\bibitem{V96}
W. Vogelsang,
%``The spin-dependent two-loop splitting functions,''
Nucl.\ Phys.\ B {475} (1996) 47; % [arXiv:hep-ph/9603366].
%%CITATION = HEP-PH 9603366;%%
% W.~Vogelsang,
%``A Rederivation of the Spin-dependent Next-to-leading Order Splitting
%Functions,''
Phys.\ Rev.\ D {54} (1996) 2023.
%  [arXiv:hep-ph/9512218].
%%CITATION = HEP-PH 9512218;%%

\bibitem{BDKM04}
A.V. Belitsky, S.E. Derkachov, G.P. Korchemsky, A.N. Manashov,
%``Quantum integrability in (super) Yang-Mills theory on the light-cone,''
Phys.\ Lett.\ B {594} (2004) 385;
%[arXiv:hep-th/0403085].
%%CITATION = HEP-TH 0403085;%%
%``Dilatation operator in (super-)Yang-Mills theories on the light-cone,''
Nucl.\ Phys.\ B {708} (2005) 115;
%  [arXiv:hep-th/0409120].
%%CITATION = HEP-TH 0409120;%%
%``Superconformal operators in Yang-Mills theories on the light-cone,''
Nucl.\ Phys.\ B {722} (2005) 191.
% [arXiv:hep-th/0503137].
%%CITATION = HEP-TH 0503137;%%

\bibitem{M83}
S. Mandelstam, Nucl. Phys. B 213 (1983) 149;
%%CITATION = NUPHA,B213,149;%%
Phys. Lett. B 121 (1983) 30;
%%CITATION = PHLTA,B121,30;%%
\\
L. Brink, O. Lindgren, B.E. Nilsson, Nucl. Phys. B 212 (1983) 401;
%%CITATION = NUPHA,B212,401;%%
Phys Lett. B 123 (1983) 323.
%%CITATION = PHLTA,B123,323;%%

\bibitem{S85}
S.J. Gates, M.T. Grisaru, M. Rocek, W. Siegel, {\em Superspace, or one thousand
and one lessons in supersymmetry}, Front. Phys. 58 (1983) 1;
%%CITATION = HEP-TH 0108200;%%
\\
M.F. Sohnius, Phys. Rept. 128 (1985) 39.
%%CITATION = PRPLC,128,39;%%

\bibitem{DO02}
F.A. Dolan, H. Osborn,
%``Superconformal symmetry, correlation functions and the operator product
%expansion,''
Nucl.\ Phys.\ B {629} (2002) 3. % [arXiv:hep-th/0112251].
%%CITATION = HEP-TH 0112251;%%

\bibitem{V97}
S. Kumano, M. Miyama,
%``Two-loop anomalous dimensions for the structure function h1,''
Phys.\ Rev.\  D {56} (1997) 2504;
% [arXiv:hep-ph/9706420].
% %%CITATION = PHRVA,D56,2504;%%
\\
W. Vogelsang,
%``Next-to-leading order evolution of transversity distributions and  Soffer's
%inequality,''
Phys.\ Rev.\ D {57} (1998) 1886; %  [arXiv:hep-ph/9706511].
%%CITATION = HEP-PH 9706511;%%
\\
A. Hayashigaki, Y. Kanazawa, Y. Koike,
%``Next-to-leading order Q**2-evolution of the transversity distribution
%h1(x,Q**2),''
Phys.\ Rev.\  D {56} (1997) 7350
%[arXiv:hep-ph/9707208].
%%CITATION = PHRVA,D56,7350;%%


\bibitem{KLOV04}
A.V. Kotikov, L.N. Lipatov, A.I. Onishchenko, V.N. Velizhanin,
%``Three-loop universal anomalous dimension of the Wilson operators in N =  4
%SUSY Yang-Mills model,''
Phys.\ Lett.\ B {595} (2004) 521 [Erratum-ibid.\ B {632} (2006) 754].
% [arXiv:hep-th/0404092].
%%CITATION = HEP-TH 0404092;%%

\bibitem{CMW91}
S. Catani, G. Marchesini, B.R. Webber,
%``QCD coherent branching and semiinclusive processes at large x,''
Nucl.\ Phys.\ B {349} (1991) 635.
%%CITATION = NUPHA,B349,635;%%

\bibitem{Gr96}
J.A. Gracey,
%``Anomalous dimension of nonsinglet Wilson operators at O (1 / N(f)) in deep
%inelastic scattering,''
Phys.\ Lett.\ B {322} (1994) 141; % [arXiv:hep-ph/9401214].
%%CITATION = HEP-PH 9401214;%%
%
%J.A. Gracey,
%``Anomalous dimensions of operators in polarized deep inelastic  scattering
%at O(1/N(f)),''
Nucl.\ Phys.\ B {480} (1996) 73.
%%CITATION = HEP-PH 9609301;%%

\bibitem{G05}
E. Gardi,
%``On the quark distribution in an on-shell heavy quark and its all-order
%relations with the perturbative fragmentation function,''
JHEP {0502} (2005) 053.
%%CITATION = HEP-PH 0501257;%%

\bibitem{Gr03}
J.A. Gracey,
%``Three loop anomalous dimension of the second moment of the transversity
%operator in the MS-bar and RI' schemes,''
Nucl.\ Phys.\ B {667} (2003) 242. %  [arXiv:hep-ph/0306163].
%%CITATION = HEP-PH 0306163;%%

\bibitem{BG97}
J.F. Bennett, J.A. Gracey,
%``Determination of the anomalous dimension of gluonic operators in deep
%inelastic scattering at O(1/N(f)),''
Nucl.\ Phys.\ B {517} (1998) 241;
%  [arXiv:hep-ph/9710364].
%%CITATION = HEP-PH 9710364;%%
%
%J.F. Bennett, J.A. Gracey,
%``Anomalous dimension of gluonic operator in polarized deep inelastic
%scattering at O(1/N(f)),''
Phys.\ Lett.\ B {432} (1998) 209. % [arXiv:hep-ph/9803446].
%%CITATION = HEP-PH 9803446;%%


\bibitem{BMN02}
D. Berenstein, J.M. Maldacena, H. Nastase, J. High Ener. Phys. 0204 (2002) 013.
%%CITATION = HEP-TH 0202021;%%

\bibitem{T03}
A.A. Tseytlin,
%``Spinning strings and AdS/CFT duality,''
{\it in} Ian Kogan Memorial Volume, `From Fields to Strings: Circumnavigating
Theoretical Physics', M. Shifman, A. Vainshtein, and J. Wheater, eds. (World
Scientific, 2004), p.1648; arXiv:hep-th/0311139.
%%CITATION = HEP-TH 0311139;%%

\bibitem{BBGK04}
A.V. Belitsky, V.M. Braun, A.S. Gorsky, G.P. Korchemsky,
  %``Integrability in QCD and beyond,''
  {\it in} Ian Kogan Memorial Volume, `From Fields to Strings: Circumnavigating
Theoretical Physics', M. Shifman, A. Vainshtein, and J. Wheater, eds. (World
Scientific, 2004), p.266;
  Int.\ J.\ Mod.\ Phys.\ A {19} (2004) 4715.
%  [arXiv:hep-th/0407232].
%%CITATION = HEP-TH 0407232;%%

\bibitem{BFST03}
N. Beisert, S. Frolov, M. Staudacher, A.A. Tseytlin, J. High Ener. Phys. 0310
(2003) 037.
%%CITATION = HEP-TH 0308117;%%

\bibitem{KZ04}
V.A. Kazakov, K. Zarembo, J. High Ener. Phys. 0410 (2004) 060.
%%CITATION = HEP-TH 0410105;%%

\bibitem{FTT06}
S. Frolov, A. Tirziu, A.A. Tseytlin,
%``Logarithmic corrections to higher twist scaling at strong coupling from
%AdS/CFT,''
arXiv:hep-th/0611269.
%%CITATION = HEP-TH 0611269;%%

\bibitem{BGK06}
A.V. Belitsky, A.S. Gorsky, G.P. Korchemsky,
%``Logarithmic scaling in gauge / string correspondence,''
Nucl.\ Phys.\ B {748} (2006) 24. % [arXiv:hep-th/0601112].
%%CITATION = HEP-TH 0601112;%%

\bibitem{SS06}
K. Sakai, Y. Satoh,
%``A large spin limit of strings on AdS(5) x S**5 in a non-compact sector,''
arXiv:hep-th/0607190.
%%CITATION = HEP-TH 0607190;%%

\bibitem{DGM97}
S.E. Derkachov, J.A. Gracey, A.N. Manashov,
%``Four loop anomalous dimensions of gradient operators in phi**4 theory,''
Eur.\ Phys.\ J.\ C {2} (1998) 569. % [arXiv:hep-ph/9705268].
%%CITATION = HEP-PH 9705268;%%

\bibitem{L93}
L.N. Lipatov, Phys. Lett. B 309 (1993) 394;
%%CITATION = PHLTA,B309,394;%%
JETP Lett. 59 (1994) 596.
%%CITATION = JTPLA,59,596;%%
%
\bibitem{FK95}
L.D. Faddeev, G.P. Korchemsky, Phys. Lett. B 342 (1995) 311.
%%CITATION = HEP-TH 9404173;%%

\bibitem{BDM98}
V.M. Braun, S.E. Derkachov, A.N. Manashov,
Phys. Rev. Lett. 81 (1998) 2020; \\
%%CITATION = HEP-PH 9805225;%%
%
V.M. Braun, S.E. Derkachov, G.P. Korchemsky, A.N. Manashov, Nucl. Phys. B 553
(1999) 355;
%%CITATION = HEP-PH 9902375;%%
\\
A.V. Belitsky, Phys. Lett. B 453 (1999) 59;
%%CITATION = HEP-PH 9902361;%%
%
Nucl. Phys. B 558 (1999) 259.
%%CITATION = HEP-PH 9903512;%%

\bibitem{L98}
L.N. Lipatov, {\sl Evolution equations in QCD}, in Perspectives in Hadronic
Physics, eds. S. Boffi, C. Ciofi Degli Atti, M. Giannini, World Scientific
(Singapore, 1998) p.\ 413;
\\
J.A. Minahan, K. Zarembo, J. High Ener. Phys.  0303 (2003) 013;
%%CITATION = HEP-TH 0212208;%%
\\
N. Beisert, M. Staudacher, Nucl. Phys. B 670 (2003) 439.
%%CITATION = HEP-TH 0307042;%%

\bibitem{K96}
G.P. Korchemsky,
%``Integrable structures and duality in high-energy {QCD},''
Nucl.\ Phys.\ B {498} (1997) 68.
%  [arXiv:hep-th/9609123].
%%CITATION = HEP-TH 9609123;%%

\bibitem{DM06}
Yu.L. Dokshitzer, G. Marchesini,
% {\em ``$\mathcal{N}=4$ SUSY Yang-Mills: three loops made simple(r)''}
arXiv:hep-th/0612248.

\end{thebibliography}
\end{document}